\documentclass{emulateapj}
\usepackage{apjfonts}

\slugcomment{Accepted for publication in ApJ}

\shorttitle{Host Galaxies of High-z AGNs in GOODS Fields}
\shortauthors{Akiyama}

\begin{document}

\title{Host Galaxies of the High-redshift AGNs in the GOODS Fields}

\author{Masayuki Akiyama\altaffilmark{1}}
\affil{Subaru Telescope, National Astronomical Observatory of Japan,
Hilo, HI, 96720}
\email{akiyama@subaru.naoj.org}

\altaffiltext{1}{
Also, Institute for Astronomy, University of Hawaii
}

\begin{abstract}
The star-formation rates and the stellar masses of the host 
galaxies of active galactic nuclei (AGNs) at high-redshifts 
are keys to understanding the evolution of the relation 
between the mass of the spheroidal component of a galaxy 
and the mass of its central black hole ($M_{\rm bulge}-M_{\bullet}$
relation). We investigate the 
host galaxies of 31 AGNs with spectroscopic redshifts between 2 and 4 
found in the deep {\it Chandra} surveys of the Great Observatories Origins 
Deep Surveys (GOODS) fields. We use the F606W, F775W, and 
F850LP band images obtained with the Advanced Camera for 
Surveys (ACS) on the Hubble Space Telescope (HST). The 
sample can be divided into 17 ``extended'' AGNs and 
14 ``compact'' AGNs based on the concentration parameter defined 
as the difference between the aperture magnitudes with 
$0.^{\prime\prime}07$  and $0.^{\prime\prime}25$ diameter. 
We derive upper limits of the UV luminosities of the host 
galaxies of the ``compact''  AGN sample, and upper and lower 
limits of the UV luminosities of the host galaxies of 
the ``extended'' AGN sample.
These limits are consistent with the knee of the luminosity 
function of the Lyman Break Galaxies (LBGs) at $z\sim3$, 
suggesting moderate star-formation rates, less than 40 $M_{\odot}$ 
yr$^{-1}$, in the host galaxies of the AGNs at $2<z_{\rm sp}<4$ 
without correcting the dust extinction. By combining the limits of
the UV luminosities with the observed $K$-band magnitudes for
the ``extended'' AGNs, we derive the upper and lower
limits of the stellar masses of their host galaxies.
The derived upper limits on the stellar masses 
range from $10^{10} M_{\odot}$ to $10^{12} M_{\odot}$.
The upper limits imply that the $M_{\rm bulge}-M_{\bullet}$
relation of the high-redshift AGNs is different from that 
of the galaxies in the nearby universe or the average 
Eddington ratio of the high-redshift AGNs is higher than 
that of low-redshift AGNs with lower-luminosity.
\end{abstract}


\section{Introduction}

Formation processes of massive black holes at the centers of galaxies have 
been one of important issues in astronomy, especially after the discovery of 
massive black holes in the centers of many massive galaxies \citep{kor95}. 
The formation processes are thought to link with the formation
of the bulges of the host galaxies. The correlations between an absolute 
magnitude of a spheroidal component of a galaxy and the mass of the 
central black hole 
(Magorrian et al. 1998; Marconi \& Hunt 2003; Haring \& Rix 2004) and
between the stellar velocity dispersion of the spheroidal component
and the black hole mass (Gebhardt et al. 2000; Merritt \& Ferrarese 2001)
suggest that the ratio between the stellar mass of the spheroidal 
component of a galaxy and the mass of its central black hole 
is constant (e.g. $M_{\bullet}/M_{\rm bulge}=0.0014$; Haring \& Rix 2004;
hereafter $M_{\rm bulge}-M_{\bullet}$ relation). 
Moreover, the similarity between the cosmic evolutions of the star formation 
rate density and of the number density of luminous quasi stellar objects 
(QSOs) suggests that the co-evolution of the stellar component and the
central black hole in a galaxy (e.g., Boyle \& Terlevich 1998; Franceschini et al. 1999). 

Active Galactic Nuclei (AGNs) are important sites of black hole growth. 
The mass density of the central black holes of galaxies 
in the nearby universe can be explained with the mass accretion 
in the observed AGNs integrated over cosmic time 
(Yu \& Tremaine 2002; Barger et al. 2001; Marconi et al. 2004).
Thus by studying the relation between the accretion parameters of AGNs and
properties of their host galaxies, we can directly reveal whether the central 
black holes and the spheroidal components of galaxies really
grow synchronously. 

For example,
by comparing the star formation rate of each host galaxy with the nuclear 
accretion rate, we can statistically connect the growth of stellar masses 
in galaxies and that of the masses of their central black holes. Discoveries of
strong far-infrared (FIR) continuum emission (e.g., Isaak et al. 2002;
Omont et al. 2002; Bertoldi et al. 2003) of high-redshift ($1.8 < z < 6.4$) 
ultra-luminous ($B$-band absolute magnitudes, $M_{B} < -26$) 
QSOs indicate that violent 
star-formation (star formation rate up to 3000 $M_{\odot}$ yr$^{-1}$) 
is associated with the QSO activity.
This means that black holes in a rapid growth phase reside in 
host galaxies under violent star-formation phase. 
By contrast, for QSOs with modest luminosity ($-25.3 < M_{B} < -21.5$),
the Hubble Space Telescope (HST) optical imaging observations of QSOs at $z\sim2$
found in the Galaxy Evolution from Morphologies and SEDs (GEMS) survey field reveal 
modest star formation in their host galaxies (Jahnke et al. 2004). They reside in 
host galaxies with object-frame UV wavelength luminosity similar to those of
Lyman Break Galaxies (LBGs) at $z\sim3$.
The corresponding star-formation rate is $2\sim16 M_{\odot}$ yr$^{-1}$ 
without correcting the dust extinction with an assumption that the UV 
emission comes from young stars under continuous star formation (Jahnke et al. 2004). 
Star-formation activity in a QSO host galaxy may 
correlate with the luminosity of the QSO (e.g., Yamada 1994).

For AGNs, the $M_{\rm bulge}-M_{\bullet}$ relation at high-redshifts
can be directly examined. If the central black hole and the spheroidal 
component of a galaxy really grow synchronously, the $M_{\rm bulge}-M_{\bullet}$ 
relation does not change with redshifts. If not, it is possible that either 
of the central black hole or the spheroidal component grow faster 
than the other in the early stage of their formation. 
In this case the $M_{\rm bulge}-M_{\bullet}$ relation at high-redshifts 
can be different from that in the nearby universe. Current observing 
facilities can not resolve the central stellar motion of high-redshift 
galaxies as high spatial resolution as for galaxies in the nearby universe, 
thus we are not able to directly measure $M_{\bullet}$ of the high-redshift 
galaxies. However, for galaxies with an AGN, the velocity width of the AGN 
broad emission line and/or luminosity of the AGN can provide us precious 
information on the mass of the central black hole in a high-redshift galaxy.

The near-infrared (NIR) imaging observations of QSOs at $z=1\sim2$ 
have been done with the NICMOS (Near Infrared Camera and Multi Object 
Spectrometer) on the HST (Kukula et al. 2001; Ridgway et al. 2001) and 
with ground-based 8-10m class telescopes with adaptive optics (AO) 
systems (Croom et al. 2004) or under good natural seeing condition 
(Falomo et al. 2004). Kukula et al. (2001) observe radio-loud and
radio-quiet QSOs with $M_{B}=-23.8 - -24.8$ mag at $z<0.5$, 
$z=1$,and $z=2$, and find that 
the object-frame $V$-band absolute magnitudes of the radio-loud 
QSOs at $z\sim1$ and $z\sim2$ are more luminous than 
low-redshift ($z<0.5$) counterpart on average and are 
consistent with the passive evolution model of a galaxy 
with $4L_V^*$ formed at $z_{\rm f}=5$. 
On the other hand, the average absolute magnitudes of host galaxies of
radio-quiet QSOs do not change significantly up to redshift 2, 
and 1 magnitude fainter than the $z_{\rm f}=5$ passive evolution 
model scaled to the absolute magnitudes of QSO host galaxies
in the nearby universe ($3L_V^*$ at $z<0.5$). Similar results
are suggested by Ridgway et al. (2001) for QSOs with $M_{B}=-22 - -24$ mag.
The upper limits of the brightnesses of host galaxies
obtained by Croom et al. (2004) for QSOs with $M_{B}=-24.5 - -27$ 
are consistent with the faint radio-quiet QSO host galaxies.
Falomo et al.(2004) observed QSOs with $M_{B}=-25.5 - -28$ at $z\sim2$ and
do not find significant evolutionary 
difference between radio-loud and radio-quiet QSOs, although 
host galaxy absolute magnitudes of radio-quiet QSOs 
are not significantly different from those by Kukula et al. (2001).  
The faint host galaxies of high-redshift radio-quiet 
QSOs imply that the stellar masses of the host galaxies are smaller 
than those of the nearby galaxies with the same black hole mass.
It is also possible that the $M_{\rm bulge}-M_{\bullet}$ 
relation does not change at high-redshifts, and
the Eddington ratio, ratio between the bolometric 
luminosity and the Eddington luminosity 
($\lambda \equiv L_{\rm bol}/L_{\rm E}$), is higher 
for high-redshift AGNs ($\lambda=1$) than for low-redshift counterparts 
($\lambda = 0.1\sim0.01$; e.g, Schade et al. 2000).
The $M_{\rm bulge}-M_{\bullet}$ relation at $z>2$
is examined with bulge velocity dispersion 
inferred from velocity width of the narrow [OIII]$\lambda$5007 
emission line and the black hole mass inferred from broad 
H$\beta$ emission line and optical luminosity (Shields et al. 2003). 
The result suggests no significant change in the relationship between 
black hole mass and bulge velocity dispersion up to redshift 3, 
although the dispersion of the distribution is large 
($\pm$1 dex; Shields et al. 2003).
These studies on the star formation rates and the
stellar masses of the host galaxies of QSOs at high-redshift
are limited in number and limited to optically-selected 
classical QSOs so far. 

In this paper, we examine star-formation rates and stellar 
masses of the host 
galaxies of X-ray-selected high-redshift AGNs in the GOODS, 
the Great Observatories Origins Deep Survey, fields, 
in order to expand the sample number of host galaxies of high-z AGNs,
and to extend the study to AGN selected in the X-ray band. 
Thanks to ultra-deep X-ray surveys with {\it Chandra} 
(Alexander et al. 2003; Giacconi et al. 2002) and intensive 
spectroscopic follow-up observations with 
8-10m class telescopes (Barger et al. 2003; Szokoly et al. 2003), 
31 AGNs are found in the redshift range $2 < z_{\rm sp} < 4$ in the
fields. The luminosities of the AGNs range from as faint as 
Seyfert galaxies (luminosity in the hard X-ray band, 
$L_{\rm 2-10keV}<10^{44}$ erg s$^{-1}$) 
to as luminous as QSOs ($L_{\rm 2-10keV}>10^{44}$ erg s$^{-1}$),
covering the most important population of AGNs whose
contribution to the Cosmic X-ray Background is significant 
(Ueda et al. 2003), 
i.e. they represent important part of the cosmic accretion history. 
Moreover, the deep high-resolution imaging observations with the
Advanced Camera for Surveys (ACS) on the HST in the 
fields \citep{gia04} enable us to 
examine the properties of the host galaxies of the high-z AGNs.

Since X-ray-selection of AGNs is less affected by the
absorption to their nucleus than optical-selection, the X-ray 
selected samples includes
obscured narrow-line AGNs as well as non-obscured broad-line AGNs.
For luminous non-obscured AGNs, due to the bright 
nuclear stellar component of the optical image, it is 
quite difficult to directly examine the properties of the 
host galaxies, especially for high-z AGNs. Even with
the HST, still the size of the point spread function 
(PSF) (FWHM is $0.^{\prime\prime}10$) is comparable  
to a 1kpc scale ($0.^{\prime\prime}12$ at $z=2$ and 
$0.^{\prime\prime}14$ at $z=4$). Additionally, the physical
sizes of high-redshift galaxies are smaller than 
those of nearby galaxies on average. Thus, we evaluate only
the upper limits of the brightnesses of the host galaxies 
in Section 3.2.

On the other hand, obscured AGNs show relatively
faint nuclear stellar component with extended host
galaxy component, thus we can examine the nature of the host
galaxies more directly than for non-obscured AGNs. 
However, it is difficult to evaluate the contribution from 
the nuclear point source to the total image in this case. 
Therefore, we evaluate the lower and upper limits on the
brightnesses of the host galaxies in Section 3.3.

We derive the constraints on the star-formation rates of 
the host galaxies from the upper and lower limits of the 
object-frame UV brightnesses.  
Moreover, using the F606W, F775W, F850LP, and $K$-bands
photometry, we derive limits on the stellar masses of the host galaxies
and discuss the relation between stellar masses of the host 
galaxies and black hole masses of the AGNs. Throughout
the paper, we use the cosmological parameters, 
$\Omega_{\rm m}=0.3$, $\Omega_{\lambda}=0.7$, and
$H_{0} = 70$ km s$^{-1}$ Mpc$^{-1}$. All magnitudes are
on the AB magnitude system.

\begin{figure*}
\begin{center}
\plotone{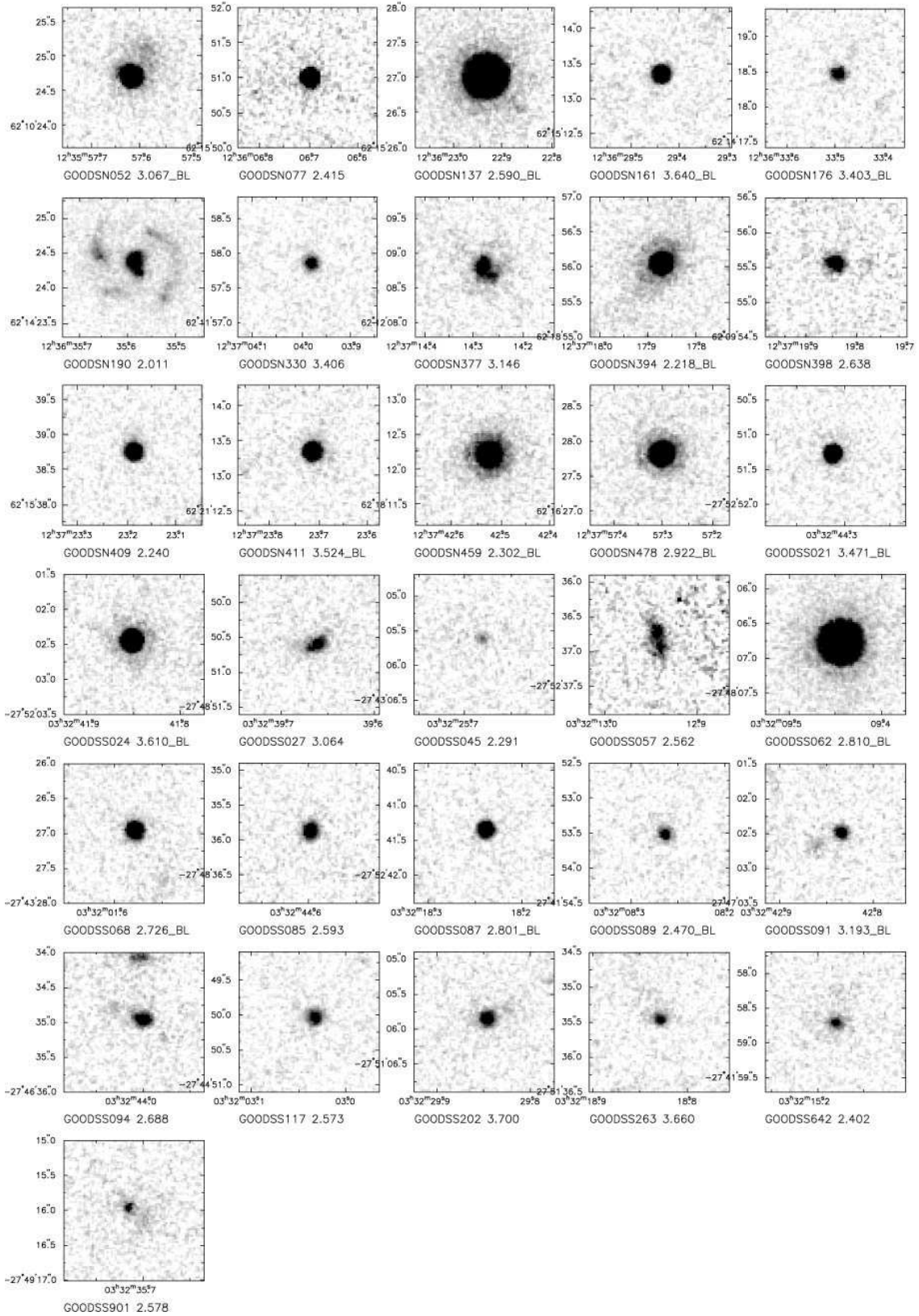}
\caption{
F775W band images of the AGNs at $2 < z_{\rm sp} < 4$ in 
the GOODS fields. Field of view is $4" \times 4"$ with north 
is up and east is left. The name and the redshift are shown in
the bottom of each panel. AGNs with a broad-emission line
in the optical spectrum are indicated with ``BL''.
\label{highzAGNi}}
\end{center}
\end{figure*}

\section{Samples and Data}

\begin{deluxetable*}{lllrrll}
\tabletypesize{\scriptsize}
\tablecaption{High Redshift AGN Sample in the GOODS Fields \label{Tab_sample}}
\tablewidth{0pt}
\tablehead{
\multicolumn{1}{c}{Name} &
\multicolumn{1}{c}{GOODS Ver.1.0 Cat. ID.} &
\multicolumn{1}{c}{$z$} &
\multicolumn{1}{c}{$F_{\rm 0.5-8keV}$\tablenotemark{a}} &
\multicolumn{1}{c}{$L_{\rm 2-10keV}$\tablenotemark{b}} &
\multicolumn{1}{c}{$K_{\rm AB}$\tablenotemark{c}} &
\multicolumn{1}{c}{Note\tablenotemark{d}} 
}
\startdata
GOODSN052 & $J123557.62+621024.7$ & 3.067 & 1.10$_{-0.14}^{+0.13}$ & 43.90$_{-0.05}^{+0.05}$ & 22.33 & BLAGN \\ 
GOODSN077 & $J123606.70+621551.0$ & 2.415 & 1.69$_{-0.11}^{+0.10}$ & 43.82$_{-0.03}^{+0.03}$ & 21.73 &  \\ 
GOODSN137 & $J123622.94+621527.0$ & 2.590 & 18.80$_{-0.32}^{+0.32}$ & 44.95$_{-0.01}^{+0.01}$ & 20.33 & BLAGN \\ 
GOODSN161 & $J123629.44+621513.3$ & 3.640 & 1.99$_{-0.11}^{+0.11}$ & 44.35$_{-0.02}^{+0.02}$ & 22.63 & BLAGN \\ 
GOODSN176 & $J123633.50+621418.4$ & 3.403 & 1.16$_{-0.09}^{+0.09}$ & 44.04$_{-0.03}^{+0.04}$ & 23.73 & BLAGN \\ 
GOODSN190 & $J123635.59+621424.3$ & 2.011\tablenotemark{e} & 2.52$_{-0.19}^{+0.18}$ & 43.79$_{-0.03}^{+0.03}$ & 21.03 &  \\ 
GOODSN330 & $J123703.99+621157.8$ & 3.406 & 0.32$_{-0.09}^{+0.08}$ & 43.48$_{-0.11}^{+0.12}$ & 23.13 &  \\ 
GOODSN377 & $J123714.28+621208.8$ & 3.146 & 0.79$_{-0.09}^{+0.08}$ & 43.79$_{-0.05}^{+0.05}$ & 21.93 &  \\ 
GOODSN394 & $J123717.88+621856.0$ & 2.218 & 9.95$_{-0.27}^{+0.26}$ & 44.50$_{-0.01}^{+0.01}$ & 21.43 & BLAGN \\ 
GOODSN398 & $J123719.84+620955.5$ & 2.638 & 3.38$_{-0.19}^{+0.18}$ & 44.22$_{-0.02}^{+0.02}$ & 22.23 &  \\ 
GOODSN409 & $J123723.19+621538.7$ & 2.240 & 1.77$_{-0.14}^{+0.13}$ & 43.76$_{-0.03}^{+0.03}$ & 22.13 &  \\ 
GOODSN411 & $J123723.72+622113.3$ & 3.524 & 1.60$_{-0.18}^{+0.17}$ & 44.22$_{-0.05}^{+0.05}$ & 22.83 & BLAGN \\ 
GOODSN459 & $J123742.53+621812.2$ & 2.302 & 20.80$_{-0.40}^{+0.39}$ & 44.86$_{-0.01}^{+0.01}$ & 20.93 & BLAGN \\ 
GOODSN478 & $J123757.31+621627.8$ & 2.922 & 4.68$_{-0.19}^{+0.19}$ & 44.48$_{-0.02}^{+0.02}$ & 20.73 & BLAGN \\ 
GOODSS021 & $J033244.31-275251.3$ & 3.471 & 1.21$_{-0.15}^{+0.14}$ & 44.08$_{-0.05}^{+0.05}$ & 23.30$^{I}$ & BLAGN \\ 
GOODSS024 & $J033241.85-275202.5$ & 3.610 & 5.73$_{-0.37}^{+0.35}$ & 44.80$_{-0.03}^{+0.03}$ & 21.31$^{I}$ & BLAGN \\ 
GOODSS027 & $J033239.67-274850.6$ & 3.064 & 7.56$_{-0.47}^{+0.45}$ & 44.74$_{-0.03}^{+0.03}$ & 21.61$^{I}$ &  \\ 
GOODSS045 & $J033225.68-274305.7$ & 2.291 & 6.12$_{-0.40}^{+0.38}$ & 44.32$_{-0.03}^{+0.03}$ & 21.73$^{I}$ &  \\ 
GOODSS057 & $J033212.94-275236.9$ & 2.562 & 5.53$_{-0.40}^{+0.37}$ & 44.41$_{-0.03}^{+0.03}$ & 22.64 &  \\ 
GOODSS062 & $J033209.45-274806.8$ & 2.810 & 7.11$_{-0.41}^{+0.39}$ & 44.62$_{-0.02}^{+0.02}$ & 19.64 & BLAGN \\ 
GOODSS068 & $J033201.58-274327.0$ & 2.726 & 8.59$_{-0.40}^{+0.38}$ & 44.67$_{-0.02}^{+0.02}$ & 22.55 & BLAGN \\ 
GOODSS085 & $J033244.60-274835.9$ & 2.593 & 1.86$_{-0.20}^{+0.18}$ & 43.95$_{-0.04}^{+0.04}$ & 23.12$^{I}$ &  \\ 
GOODSS087 & $J033218.24-275241.4$ & 2.801 & 1.00$_{-0.15}^{+0.14}$ & 43.76$_{-0.06}^{+0.07}$ & 23.32$^{I}$ & BLAGN \\ 
GOODSS089 & $J033208.27-274153.5$ & 2.470 & 1.27$_{-0.20}^{+0.18}$ & 43.73$_{-0.06}^{+0.07}$ & N.A. & BLAGN \\ 
GOODSS091 & $J033242.84-274702.5$ & 3.193 & 2.32$_{-0.21}^{+0.20}$ & 44.27$_{-0.04}^{+0.04}$ & 23.59$^{I}$ & BLAGN \\ 
GOODSS094 & $J033244.01-274635.0$ & 2.688 & 1.14$_{-0.13}^{+0.12}$ & 43.77$_{-0.05}^{+0.05}$ & 22.58$^{I}$ &  \\ 
GOODSS117 & $J033203.04-274450.1$ & 2.573 & 1.61$_{-0.18}^{+0.17}$ & 43.87$_{-0.05}^{+0.05}$ & 22.33 &  \\ 
GOODSS202 & $J033229.85-275105.9$ & 3.700 & 3.44$_{-0.34}^{+0.31}$ & 44.61$_{-0.04}^{+0.04}$ & 22.50$^{I}$ &  \\ 
GOODSS263 & $J033218.83-275135.5$ & 3.660 & 1.46$_{-0.32}^{+0.27}$ & 44.22$_{-0.09}^{+0.09}$ & 22.38$^{I}$ &  \\ 
GOODSS642 & $J033215.18-274158.7$ & 2.402 & 0.85$_{-0.15}^{+0.15}$ & 43.52$_{-0.07}^{+0.08}$ & 22.18$^{I}$ &  \\ 
GOODSS901 & $J033235.71-274916.0$ & 2.578 & 0.78$_{-0.20}^{+0.16}$ & 43.56$_{-0.10}^{+0.10}$ & 21.75$^{I}$ &  \\ 
\enddata
\tablenotetext{a}{0.5--8~keV flux from Alexandar et al. (2003) in unit of $1\times10^{-15}$
erg s$^{-1}$ cm$^{-2}$.}
\tablenotetext{b}{2--10~keV hard X-ray luminosity. It is calculated
with the 0.5--8~keV flux and the photon index of the best fit 
power-law model in the 0.5--8~keV band from Alexandar et al. (2003).}
\tablenotetext{c}{$K_{\rm AB}$ magnitude. 
For GOODSN objects, $K_{\rm AB}$ band magnitudes are derived from 
corrected-aperture $HK'_{\rm vega}$ band magnitude with $HK'_{\rm vega}-K_{\rm vega}
=0.13+0.05(I_{\rm vega}-K_{\rm vega})$ 
(Barger et al. 2003) and $K_{\rm AB} - K_{\rm vega} = 1.83$. For GOODSS objects,
the $K_{\rm AB}$-band magnitudes are mostly measured on the ISSAC
deep $K-$band image of the GOODSS field (Vandame et al. (2004) in preparation).
We use the AUTO magnitude with Sextractor in AB magnitude system. 
``I'' indicates that the magnitude is taken from the mosaiced ISSAC image.
Remaining 4 objects, $K_{\rm AB}$ magnitudes are taken from Szokoly et al.
(2003). Vega magnitude in Szokoly et al. (2003) are converted to AB magnitude 
with the conversion above. 
N.A. means $K_{\rm AB}$ magnitude is not available for the object.}
\tablenotetext{d}{BLAGN indicates an AGN with broad emission line in the
optical wavelength.}
\tablenotetext{e}{Redshift from Dawson et al. (2003)}
\end{deluxetable*}

High-redshift AGNs are selected from the catalogs 
of spectroscopically-identified optical counterparts of 
X-ray sources found in the ultra-deep {\it Chandra} surveys in the GOODS 
North (GOODSN; Barger et al. 2003) and GOODS South 
(GOODSS; Szokoly et al. 2003) fields. The flux limits of the
surveys reach $2.5\times10^{-17}$ erg s$^{-1}$ cm$^{-2}$ 
(0.5--2~keV) and $1.4\times10^{-16}$ erg s$^{-1}$ cm$^{-2}$ (2--10~keV) 
in the on-axis area of GOODSN field with 2Ms exposure time, 
and $5.2\times10^{-17}$ erg s$^{-1}$ cm$^{-2}$  (0.5--2~keV) 
and $2.8\times10^{-16}$ erg s$^{-1}$ cm$^{-2}$ (2--10~keV) in the
on-axis area of GOODSS field with 1Ms exposure time (Alexander et al. 2003).

In the GOODSN and GOODSS regions, optical images obtained with the
ACS through the F435W, F606W, F775W, and F850LP-band 
filters are available. Based on the transmission curves of the filters
and the response function of the CCD detector of the ACS, we estimate
the effective wavelengths of the F435W, F606W, F775W, and 
F850LP-band images are 4301{\AA}, 5926{\AA}, 7705{\AA}, and 9037{\AA}, 
respectively, for an object with a flat optical spectrum.

In order to escape from the incompleteness of the optical identification 
in the redshift range between $1.4<z<2.0$, i.e. ``redshift desert'', we 
set the lower redshift limit of our sample as 2. The upper redshift limit 
is determined to be 4, because for objects at below the redshifts, 
the central wavelength of the $K$-band corresponds to the wavelength 
range above 4000{\AA} break in the object frame. There are 14 (GOODSN) and 17 (GOODSS) AGNs with 
spectroscopic redshifts between $2<z_{\rm sp}<4$ in the region 
covered with the ACS imaging surveys in GOODS fields (Table 1). 
The 0.5--8~keV X-ray fluxes taken from 
Alexander et al. (2003), and 
$L_{\rm 2-10keV}$ of the objects are listed in the table. 
The $L_{\rm 2-10keV}$ is calculated from the 0.5--8~keV flux 
and the best fit photon index in the 0.5--8~keV band by 
Alexander et al. (2003). The $L_{\rm 2-10keV}$ of the sample 
ranges from $10^{43.5}$ erg s$^{-1}$ to $10^{45.0}$ erg s$^{-1}$.

The sample is not complete as X-ray flux limited AGNs
at $2<z_{\rm sp}<4$, because the optical identification of the X-ray 
sources in both of the GOODSN and GOODSS fields is not complete. 
In GOODSN (GOODSS) area, the completeness of the spectroscopic 
identification is 56\% (40\%) in total 
and 87\% (78\%) for objects brighter than $R=24$ (Barger et al. 2003;
Szokoly et al. 2003). The photometric redshift estimation by 
Barger et al. (2003) implies that there can be the same number of 
objects with only photometric redshift as the spectroscopically-identified 
objects at $2 < z < 4$. We miss some objects with fainter optical
magnitudes, which correspond to smaller star-formation rates and 
stellar masses, than those of AGNs with $z_{\rm sp}$ on average.

We examine the photometric properties of the host galaxies
of the AGNs at $2<z_{\rm sp}<4$ with the ACS GOODS Ver 0.5
images in the optical wavelength (Giavalisco et al. 2004). 
Their F775W-band images are shown in Figure 1.
The pixel scale of the ACS image is $0.^{\prime\prime}03$ pixel$^{-1}$ 
after the ``drizzle'' data reduction procedure (Fruchter \& Hook 2002). 
All of the AGNs at $2<z_{\rm sp}<4$ are detected in the GOODS Ver 1.0 catalogs
made by the GOODS team \footnote{Available from http://www.stsci.edu/science/goods/}. 
The catalogs are made by applying SExtractor object detection algorithm 
(Bertin \& Arnouts, 1996) to the F435W, F606W, F775W, and
F850LP-band images. We use all bands but F435W in this study, 
because the F435W magnitudes are affected by absorption blue-ward 
of Ly$\alpha$ for most of the AGNs at $2<z_{\rm sp}<4$.  
AUTO magnitudes, aperture magnitudes, and CLASS\_STAR stellarities of 
the objects used in the following sections are from the catalog. 
In addition to the magnitudes, a total magnitude of each AGN is determined 
by applying aperture photometry to the GOODS Ver 0.5 images. The aperture 
is determined for each object independently to cover the whole extended 
structure. The sky level is determined so as to keep the the total 
counts at large aperture radii ($\sim10$ FWHM) to a constant value. Some of the objects 
have a close companion around them, for example GOODSS091. We include 
the contribution from the companion object(s) in the photometric measurements 
below if the object is within $0.^{\prime\prime}5$ ($\sim$5kpc)
from the primary object.


We use $K_{\rm AB}$-band magnitudes of the AGNs for estimations
of the stellar masses of the host galaxies. The $K_{\rm AB}$-band 
magnitudes of the AGNs are listed in Table 1.
For all objects in GOODSN fields, the $K_{\rm AB}$-band magnitudes
are derived from corrected-aperture $HK'_{\rm vega}$ band
magnitude with $HK'_{\rm vega}-K_{\rm vega}=
0.13+0.05(I_{\rm vega}-K_{\rm vega})$(Barger et al. 1999)
and $K_{\rm AB}-K_{\rm vega}=1.83$. For 12 GOODSS AGNs, 
the $K_{\rm AB}$-band magnitudes are measured from ISAAC 
(Infrared Spectrometer And Array Camera) deep $K_s$-band 
image of the GOODSS field (Vandame et al. 2004, in preparation).
The ISAAC observations have been carried
out using the Very Large Telescope at the ESO Paranal
Observatory under Program ID:LP168.A-0485. We apply
SExtractor source detection to the images with 
a magnitude zero-point of 26.0 mag and use the AUTO magnitudes.
Other 4 GOODSS AGNs, $K_{\rm AB}$-band magnitudes
are taken from Szokoly et al. (2003). 
Vega magnitudes in Szokoly et al. (2003) are converted
to AB magnitudes with the same conversion above. 
Remaining one object (GOODSS089) is covered in neither of the data.

\section{Photometric Properties of the Host Galaxies of the High-z AGNs}

\subsection{Concentration Parameters of the Images of the High-z AGNs}

\begin{figure}
\begin{center}
\plotone{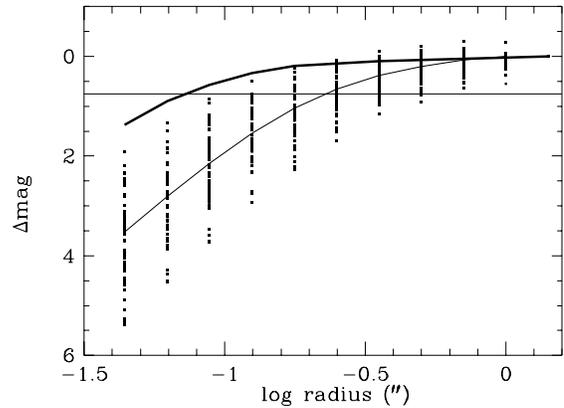}
\caption{
Average growth curves of the 400 stellar objects (thick solid line)
and 55 galaxies at $2<z_{\rm sp}<4$ (thin solid line) in the F775W band.
Growth curves of the individual galaxies at $2<z_{\rm sp}<4$ are plotted
with filled squares. All of the profiles are normalized at
$1.^{\prime\prime}4$ radius. The horizontal line 
indicates 0.75mag difference, i.e. half light.
\label{highz_prof}}
\end{center}
\end{figure}

\begin{figure}
\begin{center}
\plotone{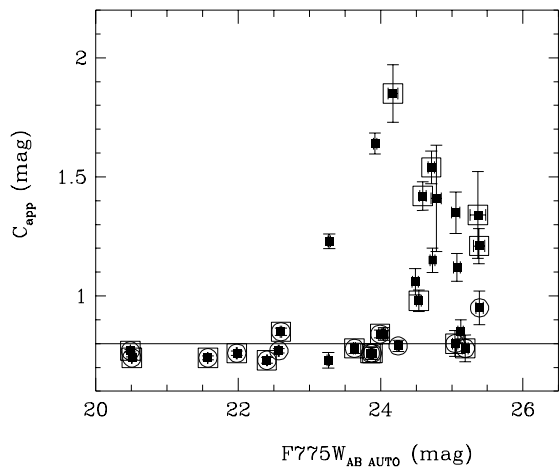}
\caption{
Concentration parameters and AUTO magnitudes
of AGNs at $2<z_{\rm sp}<4$ in the F775W band. 
The open circles indicate AGNs with a broad-line 
in their optical spectra. The ``stars'' have
$C_{\rm app}$ smaller than the stellarity limit
shown with the solid line, $C_{\rm app}=0.8$ in the band.
The large open circles and open squares indicate
AGNs with broad emission line in their optical
spectra and with $L_{\rm 2-10keV}>10^{44}$ erg s$^{-1}$,
respectively.
\label{Concent}}
\end{center}
\end{figure}

\begin{deluxetable*}{lllllll}
\tabletypesize{\scriptsize}
\tablecaption{AUTO Magnitudes and Concentration Parameters of the AGN Images \label{Tab_Cocen}}
\tablewidth{0pt}
\tablehead{
\multicolumn{1}{c}{Name} &
\multicolumn{3}{c}{Auto Mag.\tablenotemark{a}} &
\multicolumn{3}{c}{$C_{\rm app}$\tablenotemark{b}}  \\
\multicolumn{1}{c}{} &
\multicolumn{1}{c}{F606W$_{\rm AB}$} &
\multicolumn{1}{c}{F775W$_{\rm AB}$} &
\multicolumn{1}{c}{F850LP$_{\rm AB}$} &
\multicolumn{1}{c}{F606W$_{\rm AB}$} &
\multicolumn{1}{c}{F775W$_{\rm AB}$} &
\multicolumn{1}{c}{F850LP$_{\rm AB}$} 
}
\startdata
GOODSN052 & $22.71\pm0.01$ & $22.57\pm0.01$ & $22.62\pm0.01$ & $0.74\pm0.01$ & $0.77\pm0.01$ & $0.91\pm0.01$ \\ 
GOODSN077 & $23.38\pm0.01$ & $23.27\pm0.02$ & $23.13\pm0.02$ & $0.75\pm0.02$ & $0.73\pm0.03$ & $0.92\pm0.03$ \\ 
GOODSN137 & $20.49\pm0.01$ & $20.51\pm0.01$ & $20.27\pm0.01$ & $0.74\pm0.00$ & $0.74\pm0.01$ & $0.88\pm0.00$ \\ 
GOODSN161 & $24.62\pm0.02$ & $23.89\pm0.02$ & $23.70\pm0.02$ & $0.69\pm0.02$ & $0.76\pm0.03$ & $0.88\pm0.02$ \\ 
GOODSN176 & $25.26\pm0.03$ & $25.05\pm0.04$ & $25.15\pm0.05$ & $0.77\pm0.04$ & $0.80\pm0.05$ & $1.01\pm0.06$ \\ 
GOODSN190 & $23.75\pm0.01$ & $23.28\pm0.02$ & $22.91\pm0.02$ & $1.13\pm0.02$ & $1.23\pm0.03$ & $1.39\pm0.03$ \\ 
GOODSN330 & $25.30\pm0.03$ & $25.13\pm0.04$ & $25.07\pm0.04$ & $0.83\pm0.04$ & $0.85\pm0.05$ & $1.06\pm0.05$ \\ 
GOODSN377 & $24.31\pm0.02$ & $23.92\pm0.03$ & $23.57\pm0.02$ & $1.68\pm0.03$ & $1.64\pm0.04$ & $1.71\pm0.04$ \\ 
GOODSN394 & $22.83\pm0.01$ & $22.60\pm0.01$ & $22.29\pm0.01$ & $0.85\pm0.01$ & $0.85\pm0.02$ & $0.98\pm0.01$ \\ 
GOODSN398 & $24.84\pm0.03$ & $24.59\pm0.05$ & $24.42\pm0.05$ & $1.33\pm0.04$ & $1.42\pm0.06$ & $1.57\pm0.06$ \\ 
GOODSN409 & $24.48\pm0.01$ & $24.04\pm0.02$ & $23.74\pm0.02$ & $0.84\pm0.02$ & $0.84\pm0.03$ & $0.99\pm0.02$ \\ 
GOODSN411 & $23.96\pm0.01$ & $23.63\pm0.01$ & $23.62\pm0.02$ & $0.73\pm0.02$ & $0.78\pm0.02$ & $0.96\pm0.02$ \\ 
GOODSN459 & $21.62\pm0.01$ & $21.58\pm0.01$ & $21.31\pm0.01$ & $0.70\pm0.01$ & $0.74\pm0.01$ & $0.87\pm0.01$ \\ 
GOODSN478 & $22.47\pm0.01$ & $21.98\pm0.01$ & $21.87\pm0.01$ & $0.73\pm0.01$ & $0.76\pm0.01$ & $0.91\pm0.01$ \\ 
GOODSS021 & $23.99\pm0.01$ & $23.86\pm0.02$ & $23.98\pm0.02$ & $0.71\pm0.02$ & $0.76\pm0.03$ & $0.89\pm0.03$ \\ 
GOODSS024 & $22.89\pm0.01$ & $22.40\pm0.01$ & $22.43\pm0.01$ & $0.68\pm0.01$ & $0.73\pm0.01$ & $0.84\pm0.01$ \\ 
GOODSS027 & $25.53\pm0.05$ & $24.71\pm0.05$ & $24.55\pm0.05$ & $1.46\pm0.07$ & $1.54\pm0.07$ & $1.66\pm0.07$ \\ 
GOODSS045 & $25.87\pm0.09$ & $25.37\pm0.11$ & $24.83\pm0.08$ & $1.41\pm0.15$ & $1.34\pm0.18$ & $1.84\pm0.17$ \\ 
GOODSS057 & $24.13\pm0.03$ & $24.17\pm0.07$ & $23.76\pm0.19$ & $1.84\pm0.07$ & $1.85\pm0.12$ & $1.96\pm0.17$ \\ 
GOODSS062 & $20.80\pm0.01$ & $20.49\pm0.01$ & $20.31\pm0.01$ & $0.72\pm0.00$ & $0.77\pm0.00$ & $0.89\pm0.00$ \\ 
GOODSS068 & $24.14\pm0.01$ & $23.99\pm0.01$ & $23.78\pm0.02$ & $0.79\pm0.01$ & $0.84\pm0.02$ & $0.95\pm0.02$ \\ 
GOODSS085 & $25.16\pm0.02$ & $24.73\pm0.03$ & $24.61\pm0.03$ & $1.05\pm0.04$ & $1.15\pm0.05$ & $1.32\pm0.05$ \\ 
GOODSS087 & $24.37\pm0.01$ & $24.25\pm0.02$ & $24.28\pm0.02$ & $0.73\pm0.02$ & $0.79\pm0.02$ & $0.93\pm0.03$ \\ 
GOODSS089 & $25.49\pm0.03$ & $25.39\pm0.05$ & $25.14\pm0.05$ & $0.84\pm0.04$ & $0.95\pm0.07$ & $1.08\pm0.07$ \\ 
GOODSS091 & $25.04\pm0.02$ & $25.19\pm0.04$ & $25.18\pm0.04$ & $0.75\pm0.03$ & $0.78\pm0.06$ & $0.89\pm0.06$ \\ 
GOODSS094 & $24.64\pm0.02$ & $24.49\pm0.04$ & $24.37\pm0.04$ & $1.07\pm0.04$ & $1.06\pm0.05$ & $1.23\pm0.06$ \\ 
GOODSS117 & $25.80\pm0.05$ & $25.07\pm0.05$ & $24.70\pm0.04$ & $1.13\pm0.06$ & $1.12\pm0.06$ & $1.23\pm0.05$ \\ 
GOODSS202 & $25.18\pm0.03$ & $24.54\pm0.03$ & $24.64\pm0.05$ & $0.80\pm0.04$ & $0.98\pm0.05$ & $1.34\pm0.06$ \\ 
GOODSS263 & $25.67\pm0.05$ & $25.38\pm0.08$ & $25.09\pm0.08$ & $1.24\pm0.06$ & $1.21\pm0.07$ & $1.67\pm0.10$ \\ 
GOODSS642 & $25.60\pm0.05$ & $25.06\pm0.06$ & $24.85\pm0.06$ & $1.25\pm0.07$ & $1.35\pm0.09$ & $1.68\pm0.08$ \\ 
GOODSS901 & $25.37\pm0.06$ & $24.78\pm0.07$ & $24.66\pm0.08$ & $1.01\pm0.15$ & $1.41\pm0.22$ & $1.49\pm0.23$ \\ 
\enddata
\tablenotetext{a}{AUTO magnitude from GOODS catalog version 1.0.}
\tablenotetext{b}{Concentration parameter derived with 
$C_{\rm app}\equiv{\rm mag}_{\rm app}(<0.^{\prime\prime}07)-{\rm mag}_{\rm app}(<0.^{\prime\prime}25)$.}
\end{deluxetable*}

We examine the concentration parameters of the images
of each object in F606W, F775W, and F850LP bands, in 
order to divide the high-z AGNs into two samples; AGNs dominated
by PSF component without a significant extended component 
(``compact'' AGNs), and AGNs with an extended component
(``extended'' AGNs). We adopt a concentration parameter 
derived from the difference of aperture magnitudes at two 
radii, following Abraham et al. (1994). We choose the
inner radius so as to match the size of the PSF and outer
radius to match the typical size of galaxies at $2<z_{\rm sp}<4$
the same redshift range as the AGNs. 

The radial profile of the PSF is determined by
the average profile of ``stars'' selected in the 
following way from the ACS images. First, we choose 
objects whose SExtractor stellarity parameters, CLASS\_STAR, 
in all of the F606W, F775W, and F850LP bands are larger than 0.95
in the magnitude range $21<m_{\rm F775W}<26$ mag. In
the magnitude range ``stars'' form a clear sequence with CLASS\_STAR
around 0.98 in the magnitude vs. CLASS\_STAR diagram, thus 
``stars'' can be clearly separated from extended objects using 
the threshold, which enclose the outer envelop of the sequence.
The magnitude range covers slightly fainter than that of 
the $2<z_{\rm sp}<4$ AGN sample (i.e. $20.5<m_{\rm F775W}<25.4$).
Then, we check the images of the ``stars'' by eye and remove
objects affected by nearby bright objects, edge of the field 
of view, and residual of bad pixels and cosmic-rays. 
About 20\% of
the objects are rejected, and 205 and 195 stars are remained in 
the GOODSN and the GOODSS fields, respectively.

The profiles of galaxies are examined using the
galaxies at $2 < z_{\rm sp} < 4$ selected from the compiled 
redshift survey catalogs in the GOODSN field (Cohen et al. 2000; 
Cowie et al. 2004). There are 55 galaxies in the redshift 
range within the area covered by the ACS imaging, excluding
the objects detected by {\it Chandra}. The magnitudes
of the 55 galaxies are distributed between $22 < m_{\rm F775W} < 27$ mag.

The half-light radii of the average profiles
are measured to be $0.^{\prime\prime}08$ (2.6 pixel) for 
the ``stars'' and $0.^{\prime\prime}23$ (7.7 pixel) for 
the $2 < z_{\rm sp} < 4$ galaxies from the average growth
curves shown in Figure~\ref{highz_prof} with thick solid
line and thin solid line, respectively. The half-light
radius is consistent with that measured in 
the bright U-dropout LBG sample (Bouwens, Broadhurst, \& Illingworth 2003).
Based on the results, from aperture magnitudes available 
in the GOODS Ver. 1.0 catalog, we choose those with radii 
of $0^{\prime\prime}.07$  (2.08 pixel) 
and $0^{\prime\prime}.25$ (8.34 pixel), 
and define the concentration parameter as 
$C_{\rm app} = {\rm mag}_{\rm app}(<0^{\prime\prime}.07) 
- {\rm mag}_{\rm app}(<0^{\prime\prime}.25)$. It should
be noted that a stellar object has a smaller $C_{\rm app}$
than an extended object. We use the concentration
parameter instead of the CLASS\_STAR, which is derived by 
a neural network method with isophotal areas, to evaluate the 
stellarity of an object. Because we can more simply estimate 
the uncertainty for the
stellarity using the concentration parameter than using the 
CLASS\_STAR.

The distribution of $C_{\rm app}$ of the ``stars''
does not change significantly in the magnitude range of the
``stars''. Based on the cumulative distributions of $C_{\rm app}$,
we decide to regard an object with the $C_{\rm app}$ smaller than 
0.79, 0.80, and 0.96 in the F606W, F775W, 
and F850LP bands, respectively, as a stellar object. 
Ninety-five percent of the ``stars'' fall below the
stellarity limits. The $C_{\rm app}$ of the AGNs at $2<z_{\rm sp}<4$
are summarized in Table 2 along with the AUTO magnitudes.

Based on the $C_{\rm app}$ of the F606W, F775W, and
F850LP images, we divide the sample of the AGNs 
at $2<z_{\rm sp}<4$ into ``extended'' and ``compact'' AGNs.
Seventeen $2<z_{\rm sp}<4$ AGNs have concentration parameters 
larger than the stellarity limits in all of the F606W, F775W, and 
F850LP band images. Hereafter, we call them ``extended'' AGNs. 
Remaining 14 AGNs are not significantly extended at least 
in one of the three bands (``compact'' AGN). Because we
use the criteria that the ``extended'' AGNs are significantly
extended in all of the three bands, we expect the number
of the contamination of purely point-like source to the 
``extended'' sample is negligible. On the contrary,
the ``compact'' sample can have objects with faint
extended components in one band.

In Figure~\ref{Concent}, the F775W band $C_{\rm app}$ of 
the AGNs are plotted against their F775W band AUTO magnitudes.
The large open circles in the figure indicate the objects
that show a broad-emission line in their optical spectra. 
They tend to have smaller $C_{\rm app}$, 
which means they show a stellar image. The large open squares indicate
AGNs with $L_{\rm X}>10^{44}$ erg s$^{-1}$. There is no clear
correlation between the hard X-ray luminosity of the AGN and the $C_{\rm app}$.

\subsection{Upper Limits on the Brightnesses of the Host Galaxies 
of the ``Compact'' AGNs}

\begin{figure}
\begin{center}
\plotone{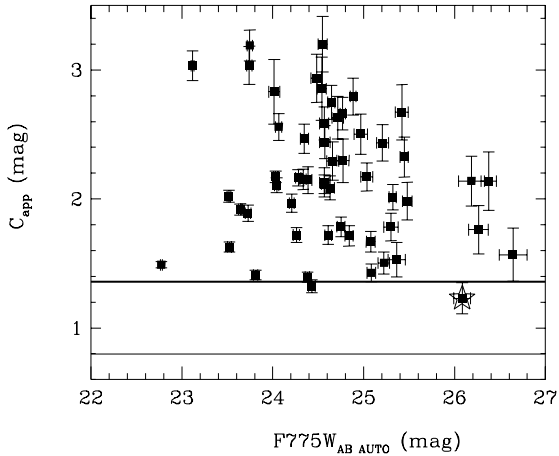}
\caption{
Concentration parameters and AUTO magnitudes
of the 55 $2<z_{\rm sp}<4$ galaxies in the F775W band. 
The thick solid line represents the $C_{\rm app}$
value of a $z=3$ object with the de Vaucouleurs profile
with effective radius of 1 kpc and axis ratio of 0.4,
which represents the most concentrated host galaxies 
observed in the nearby universe (e.g., Schade et al. 2000).
The solid line shows the stellarity limit.
The open star indicates the most concentrated faint
object used as the upper limit for GOODSN176 and GOODSS091,
see Section 3.2 for details.
\label{highz_capp}}
\end{center}
\end{figure}

\begin{figure*}
\begin{center}
\plotone{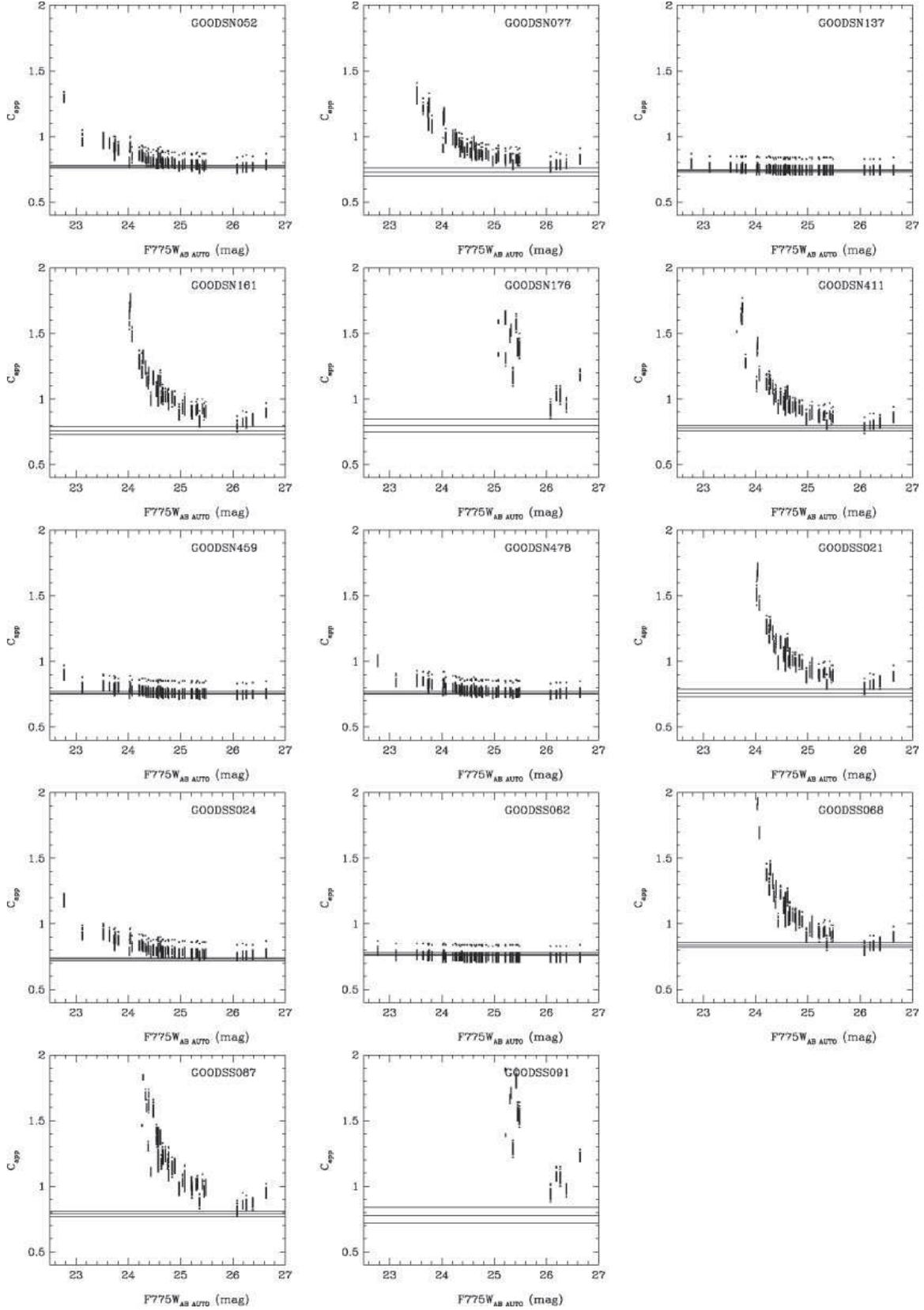}
\caption{
The distributions of the $C_{\rm app}$ 
parameters of the simulated images and the
magnitudes of the model galaxy component used
in the simulated images. A dot represent
a result with a simulated image. Each panel 
corresponds to an ``compact'' AGN.
Horizontal lines in each panel represent
the 1$\sigma$ range of the $C_{\rm app}$ of the AGN. 
\label{PSF_LBGi}}
\end{center}
\end{figure*}

Due to the bright nuclear stellar components of the 
``compact'' AGNs, it is impossible to directly estimate the
brightnesses of their host galaxies. We can only determine the upper 
limits of the brightnesses of the host galaxies of the ``compact'' AGNs. 
We compare the $C_{\rm app}$ of the AGNs with those of 
simulated images, which consist of a stellar nuclear component 
and an extended host galaxy component. If the stellar component
dominates a simulated image, the $C_{\rm app}$ of the simulated image 
should be close to that of the image of a ``compact'' AGN. The
contribution from the extended component increases, the
$C_{\rm app}$ of the simulated images increases, and at some 
points, the $C_{\rm app}$ exceeds the observed upper limits.
We use the brightness of the extended
component with which the $C_{\rm app}$ of the simulated image
exceed the observed $C_{\rm app}$ value as the upper limit
of the brightness of the host galaxy. We assume that the nuclear
component comes from the AGN component, not from a nuclear 
star-burst component.

The simulated images are constructed by adding the images of stars to 
the images of high-redshift galaxies. Because we do not have information 
on the morphological distribution of high-redshift galaxies, we do
not make model galaxy image with typical galaxy profiles, such as
the de Vaucouleurs profile, observed in the nearby universe. As the model 
host galaxy images, we use the images of high-redshift field (non-AGN) 
galaxies in the GOODS fields themselves, i.e. the 55 $2<z_{\rm sp}<4$ 
galaxies selected in Section 3.1. It should be noted that the upper 
limit evaluation can be affected by morphological distribution of 
the model host galaxies. If the host galaxies of the AGNs are 
systematically more concentrated than the field galaxies, the 
estimated upper limits can be fainter than the real host galaxy 
brightness. The magnitudes and the concentration parameters of 
the 55 $2 < z_{\rm sp} < 4$ galaxy are shown in Figure~\ref{highz_capp}.
They cover quite large range of the $C_{\rm app}$ value,
and the most concentrated objects are as concentrated as
an $z=3$ object with the de Vaucouleurs profile with effective 
radius of 1 kpc and axis ratio of 0.4 ($C_{\rm app}$ of 1.47).
The profile represents the most concentrated
host galaxies observed in the nearby universe (e.g., Shade et al. 2000).
Thus, the assumption that the $2<z_{\rm sp}<4$ galaxy sample
covers the morphologies of the most concentrated AGN host galaxies
is acceptable.
 
For the ``star'' sample, we select from 
from the 400 ``stars'' used in Section 3.1. 
We only use ``stars'' in the magnitude range between the magnitude 
of the brightest ``compact'' AGN (GOODSN137) and one magnitude 
fainter than that for the simulation.
There are 19, 38, and 53 ``stars'' in the magnitude range
in the F606W, F775W, and F850LP bands. We normalize the 
image of a star so that the total magnitude of the 
simulated image match the observed magnitude of each AGN.
For each ``compact'' AGN, $19\times55=1045$ (F606W),
$38\times55=2090$ (F775W), and $53\times55=2915$ (F850LP) 
simulated images are typically made. 
We do not change the normalization of the model galaxy image,
in order to escape from changing of the signal-to-noise ratio
of the simulated images. Thus, for the two faint ``compact''
AGNs, GOODSN176 and GOODSS091, only $2<z_{\rm sp}<4$ 
galaxies fainter than the AGNs are used for the simulation.

The distribution of the measured $C_{\rm app}$ of the simulated
images and the magnitudes of the model galaxy components used
in the image is shown in Figure~\ref{PSF_LBGi} for the F775W-band case. 
Each dot represents 
a simulated image. A simulated image made with
a brighter model galaxy component have larger $C_{\rm app}$ on average 
than that with a fainter galaxy component as expected.
The scatter in the vertical direction at a 
certain magnitude is owing to the difference of the
``star'' images used in the simulations. The scatter represents
the effect of the PSF variation in the field of view.

The horizontal solid lines in each panel indicate
the concentration parameter of each AGN and 1 $\sigma$ range. 
We use the magnitude of the brightest $2<z_{\rm sp}<4$
galaxy with which the simulated image with the smallest
concentration parameter exceed the 1$\sigma$ upper limit of 
the observed $C_{\rm app}$ value as the upper limits of the
brightnesses of the host galaxies.  
The resulting upper limits on the brightnesses of the host galaxies 
are summarized in ``upper limit'' column of Table~\ref{Tab_limits1}.
For GOODSN176 and GOODSS091, even the simulated image with
the faintest model galaxy exceeds their upper limits of $C_{\rm app}$.
We use the magnitude of the most concentrated faint object in 
the 55 $2<z_{\rm sp}<4$ galaxies (indicated with open star in 
Figure~\ref{highz_capp}) as their upper limits.
The nucleus of GOODSN137 and GOODSS062 are too bright to 
determine useful upper limits of the brightnesses of their
host galaxies. We remove the two objects from discussions below.

We expect that the uncertainty on the upper limits are less than
0.5 mag, based on the number of the most concentrated galaxies
in the $2<z_{\rm sp}<4$ galaxies (about 1 per 0.5 mag bin, see Figure~\ref{highz_capp}).
Because we do not know the morphological distribution of the
AGN host galaxies at high redshifts, especially as a function of
magnitude, it is impossible to rigidly estimate the uncertainty of the 
upper limits. 

\subsection{Upper and Lower Limits on the Brightness of the 
Host Galaxies of the ``Extended'' AGNs}

Even for the ``extended'' AGNs, it is still difficult to 
directly evaluate the brightness of the host galaxy by 
fitting the nuclear and the host galaxy components to the
2-dimensional images of the AGN. Therefore, we robustly 
estimate the upper and lower limits of their brightness 
as follows. 

We use the total magnitudes of the ``extended'' AGNs as the 
upper limits of the brightnesses of the host galaxies.
The total magnitude of an AGN include the host galaxy 
component and the nuclear component, thus the magnitude
of the host galaxy has to be fainter than the total magnitude.
The total magnitudes of the ``extended'' AGNs are listed
in Table~\ref{Tab_limits2}.

We evaluate the lower limit of the brightness of the host 
galaxy by subtracting the maximal PSF contribution whose 
central surface brightness matches the peak surface 
brightness of the total image. Because the central surface 
brightness of the nuclear component can not exceed the 
peak surface brightness of the total image.
We use the aperture magnitudes within a $0.^{\prime\prime}04$ 
(1.46 pixel) radius as the peak surface brightness. 
The average profile of the ``stars'' (Figure~\ref{highz_prof})
is used as the average PSF radial profile.

The profile of each ``star'' can be different from the
average profile of the ``stars'', due to a PSF variation
across the field of view and an insufficient signal-to-noise
ratio. For the 400 ``stars'' selected in Section 3.1, 
the total magnitude estimated from the aperture magnitude 
is consistent with the measured total magnitude within 
$\pm 0.2$ mag with systematic offset less than 0.05 mag. 
Therefore for the ``extended'' AGNs, we can regard the 
magnitude difference over 0.2 mag as the significant 
difference between the total magnitude and the maximal
PSF contribution of the nuclear component.

\begin{figure}
\plotone{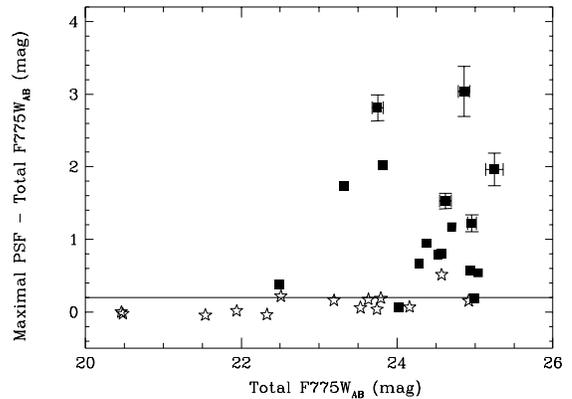}
\caption{
Differences between the total F775W-band magnitudes and the maximal 
PSF contributions in the band of ``extended'' (filled square) and 
the ``compact'' (open star) AGNs as a function of the total magnitudes.
Only error bar that is significantly larger than each point is shown.
The horizontal solid line show the upper envelop of the
distribution of the differences of the ``stars''.  
\label{PSF_APP}}
\end{figure}

The magnitude differences between the total magnitudes
and the maximal PSF contributions for the ``extended'' AGNs are shown 
in Figure~\ref{PSF_APP} with filled squares as a function of 
the total magnitudes. All of the ``extended'' AGNs, except for
GOODSN330 and GOODSN409, have magnitude difference larger than 0.2 mag. 
We evaluate
lower limit of the brightness of the host galaxy with the 
residual flux after subtracting the maximal PSF contribution 
from the total magnitude. The resulting
lower limits are shown in column ``Lower Limits'' in 
Table~\ref{Tab_limits2}. 

\begin{deluxetable*}{lllllll}
\tabletypesize{\scriptsize}
\tablecaption{Upper and Lower Limits on the Brightnesses of the Host Galaxies 
of the ``Extended'' AGNs \label{Tab_limits2}}
\tablewidth{0pt}
\tablehead{
\multicolumn{1}{c}{Name} &
\multicolumn{3}{c}{Total Mag. (Upper Limit)\tablenotemark{a}} &
\multicolumn{3}{c}{Lower Limit\tablenotemark{b}} \\
\multicolumn{1}{c}{} &
\multicolumn{1}{c}{F606W$_{\rm AB}$} &
\multicolumn{1}{c}{F775W$_{\rm AB}$} &
\multicolumn{1}{c}{F850LP$_{\rm AB}$} &
\multicolumn{1}{c}{F606W$_{\rm AB}$} &
\multicolumn{1}{c}{F775W$_{\rm AB}$} &
\multicolumn{1}{c}{F850LP$_{\rm AB}$} 
}
\startdata
GOODSN190 & $23.70\pm0.01$ & $23.31\pm0.02$ & $22.98\pm0.01$ & $23.99_{-0.05}^{+0.07}$ & $23.56_{-0.05}^{+0.06}$ & $23.23_{-0.05}^{+0.06}$ \\
GOODSN330 & $25.14\pm0.03$ & $24.99\pm0.04$ & $24.94\pm0.04$ & $26.84_{-0.54}^{+1.59}$ & $26.98_{-0.69}^{+9.99}$ & $27.13_{-0.80}^{+9.99}$ \\
GOODSN377 & $24.26\pm0.02$ & $23.82\pm0.03$ & $23.54\pm0.02$ & $24.45_{-0.05}^{+0.05}$ & $24.00_{-0.07}^{+0.07}$ & $23.76_{-0.06}^{+0.06}$ \\
GOODSN394 & $22.63\pm0.01$ & $22.48\pm0.01$ & $22.18\pm0.01$ & $23.77_{-0.30}^{+0.51}$ & $23.82_{-0.37}^{+0.73}$ & $23.90_{-0.55}^{+1.67}$ \\
GOODSN398 & $24.69\pm0.03$ & $24.57\pm0.04$ & $24.03\pm0.05$ & $25.34_{-0.14}^{+0.20}$ & $25.27_{-0.15}^{+0.22}$ & $24.49_{-0.09}^{+0.12}$ \\
GOODSN409 & $24.31\pm0.01$ & $24.02\pm0.02$ & $23.64\pm0.02$ & $26.09_{-0.58}^{+2.02}$ & $27.04_{-1.38}^{+9.99}$ & $26.62_{-1.34}^{+9.99}$ \\
GOODSS027 & $25.49\pm0.05$ & $24.70\pm0.05$ & $24.45\pm0.05$ & $25.85_{-0.10}^{+0.10}$ & $25.15_{-0.09}^{+0.12}$ & $24.89_{-0.09}^{+0.12}$ \\
GOODSS045 & $25.82\pm0.09$ & $25.25\pm0.11$ & $24.76\pm0.08$ & $26.01_{-0.18}^{+0.18}$ & $25.44_{-0.22}^{+0.22}$ & $24.92_{-0.21}^{+0.21}$ \\
GOODSS057 & $23.89\pm0.03$ & $23.75\pm0.07$ & $23.84\pm0.19$ & $23.98_{-0.10}^{+0.10}$ & $23.83_{-0.18}^{+0.18}$ & $23.95_{-0.30}^{+0.30}$ \\
GOODSS085 & $25.13\pm0.02$ & $24.52\pm0.03$ & $24.46\pm0.03$ & $26.15_{-0.25}^{+0.41}$ & $25.24_{-0.16}^{+0.23}$ & $25.30_{-0.19}^{+0.29}$ \\
GOODSS089 & $25.41\pm0.03$ & $25.04\pm0.05$ & $25.01\pm0.05$ & $27.63_{-0.82}^{+9.99}$ & $26.06_{-0.25}^{+0.41}$ & $26.72_{-0.54}^{+1.63}$ \\
GOODSS094 & $24.54\pm0.02$ & $24.38\pm0.04$ & $24.24\pm0.04$ & $25.10_{-0.12}^{+0.16}$ & $24.97_{-0.12}^{+0.17}$ & $24.74_{-0.10}^{+0.13}$ \\
GOODSS117 & $25.49\pm0.05$ & $24.94\pm0.05$ & $24.42\pm0.04$ & $26.23_{-0.17}^{+0.24}$ & $25.91_{-0.24}^{+0.38}$ & $25.32_{-0.21}^{+0.33}$ \\
GOODSS202 & $25.11\pm0.03$ & $24.28\pm0.03$ & $24.59\pm0.05$ & $26.91_{-0.59}^{+2.16}$ & $25.13_{-0.20}^{+0.30}$ & $25.45_{-0.20}^{+0.30}$ \\
GOODSS263 & $25.28\pm0.05$ & $24.62\pm0.08$ & $24.92\pm0.08$ & $25.67_{-0.08}^{+0.10}$ & $24.92_{-0.11}^{+0.11}$ & $25.30_{-0.13}^{+0.13}$ \\
GOODSS642 & $25.47\pm0.05$ & $24.96\pm0.06$ & $24.83\pm0.06$ & $25.91_{-0.10}^{+0.12}$ & $25.39_{-0.11}^{+0.11}$ & $25.38_{-0.11}^{+0.16}$ \\
GOODSS901 & $25.04\pm0.06$ & $24.86\pm0.07$ & $24.48\pm0.08$ & $25.10_{-0.21}^{+0.21}$ & $24.93_{-0.34}^{+0.34}$ & $24.53_{-0.40}^{+0.40}$ \\
\enddata
\tablenotetext{a}{Total magnitude of the AGN derived from aperture photometry. See Section 2 for
the details.}
\tablenotetext{b}{The difference between the total magnitude and the maximal PSF 
contribution. $+9.99$ in the uncertainty of the lower limit represents no lower uncertainty is
derived for the lower limit. See Section 3.3 for the details.}
\end{deluxetable*}

The uncertainty of the lower limit 
consists of the uncertainties of the $0.^{\prime\prime}04$ aperture
magnitude, the total magnitude, and the profile of the PSF, 
i.e. conversion factor from the $0.^{\prime\prime}04$ aperture magnitude
to the maximal PSF contribution. The $\pm$0.2 mag
dispersion in the conversion factor mentioned in the previous 
paragraph dominates the uncertainty
of the lower limits estimations in most of the cases.
The estimated uncertainties are listed in Table~\ref{Tab_limits2}.
For GOODSN330 and GOODSN409, the magnitude differences between
the total magnitudes and the maximal PSF contributions are
less than 0.2 mag, thus we cannot determine their lower uncertainty
of the lower limits (shown with $+9.99$ in Table~\ref{Tab_limits2}).
  
For comparison, the ``compact'' AGNs are also plotted in the 
same figure with the star marks. All but GOODSN176 have the magnitude 
difference of less than 0.2 mag which is consistent with being the
stellar image. GOODSN176 has magnitude difference of 0.51 mag
in the F775W-band and consistent with the $C_{\rm app}$
value of 0.80 which is close to the stellarity threshold value 
in the band. Because the object has $C_{\rm app}$ value less than 
stellarity threshold in the F606W-band, it is classified 
as a ``compact'' AGN.

\begin{deluxetable*}{lcccccc}
\tabletypesize{\scriptsize}
\tablecaption{Total Magnitudes and Upper Limits on the Brightnesses of the Host Galaxies of the ``Compact'' AGNs \label{Tab_limits1}}
\tablewidth{0pt}
\tablehead{
\multicolumn{1}{c}{Name} &
\multicolumn{3}{c}{Total Mag.\tablenotemark{a}} &
\multicolumn{3}{c}{Upper Limit\tablenotemark{b}} \\
\multicolumn{1}{c}{} &
\multicolumn{1}{c}{F606W$_{\rm AB}$} &
\multicolumn{1}{c}{F775W$_{\rm AB}$} &
\multicolumn{1}{c}{F850LP$_{\rm AB}$} &
\multicolumn{1}{c}{F606W$_{\rm AB}$} &
\multicolumn{1}{c}{F775W$_{\rm AB}$} &
\multicolumn{1}{c}{F850LP$_{\rm AB}$} 
}
\startdata
GOODSN052 & $22.64\pm0.00$ & $22.51\pm0.01$ & $22.50\pm0.01$ & 24.3 & 24.0 & 22.5 \\
GOODSN077 & $23.34\pm0.01$ & $23.19\pm0.02$ & $23.04\pm0.02$ & 24.9 & 25.4 & 24.5 \\
GOODSN137 & $20.46\pm0.00$ & $20.48\pm0.00$ & $20.22\pm0.00$ & $\cdots$ & $\cdots$ & $\cdots$ \\
GOODSN161 & $24.49\pm0.02$ & $23.63\pm0.02$ & $23.48\pm0.02$ & 26.2 & 25.4 & 26.1 \\
GOODSN176 & $25.13\pm0.02$ & $24.57\pm0.04$ & $24.97\pm0.05$ & 26.2 & 26.1 & 26.1 \\
GOODSN411 & $23.91\pm0.01$ & $23.53\pm0.01$ & $23.45\pm0.02$ & 26.6 & 25.4 & 24.9 \\
GOODSN459 & $21.57\pm0.00$ & $21.54\pm0.00$ & $21.24\pm0.00$ & 23.9 & 23.7 & 23.9 \\
GOODSN478 & $22.41\pm0.00$ & $21.94\pm0.01$ & $21.79\pm0.01$ & 24.3 & 23.7 & 23.7 \\
GOODSS021 & $23.92\pm0.01$ & $23.74\pm0.02$ & $23.88\pm0.02$ & 26.2 & 25.4 & 26.1 \\
GOODSS024 & $22.83\pm0.01$ & $22.33\pm0.01$ & $22.35\pm0.01$ & 26.2 & 25.0 & 26.1 \\
GOODSS062 & $20.77\pm0.00$ & $20.46\pm0.00$ & $20.26\pm0.00$ & $\cdots$ & 22.8 & 22.8 \\
GOODSS068 & $23.95\pm0.01$ & $23.79\pm0.01$ & $23.55\pm0.01$ & 25.6 & 25.0 & 25.3 \\
GOODSS087 & $24.19\pm0.01$ & $24.16\pm0.02$ & $24.14\pm0.02$ & 26.2 & 26.1 & 26.1 \\
GOODSS091 & $24.92\pm0.02$ & $24.92\pm0.04$ & $25.07\pm0.04$ & 26.2 & 26.1 & 26.1 \\
\enddata
\tablenotetext{a}{Total magnitude of the AGN derived from aperture photometry. See Section 2
for the details.}
\tablenotetext{b}{Upper limits of the brightness of the host galaxies. See Section 3.2 for
the details.}
\end{deluxetable*}

\section{Results}

\subsection{Absolute UV Magnitudes of the Host Galaxies}

\begin{figure*}
\begin{center}
\plotone{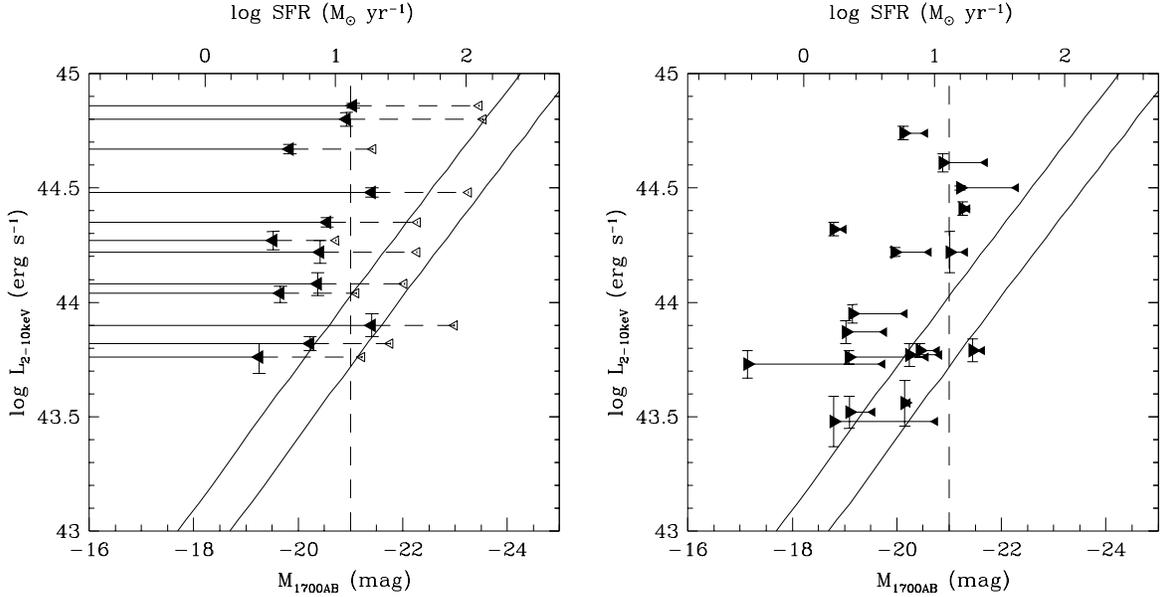}
\caption{
Left) Upper limits on the UV luminosity $M_{\rm AB 1700}$ of 
the host galaxies of the ``compact'' AGNs (filled triangle) 
along with the total UV absolute magnitudes (open triangle). 
Right) Upper (small filled triangle) and lower (large filled triangle)
limits on the UV luminosity of the host galaxies of the 
``extended'' AGNs. In both panels, the horizontal solid lines 
represent the UV luminosity range allowed. 
The horizontal dashed lines in the left panel connect the
upper limit with the total magnitude. The vertical dashed line 
indicates the knee of the UV luminosity function of $z\sim3$ LBGs 
(Steidel et al. 1999). The corresponding star-formation rate to the 
UV luminosity is shown in top horizontal axis.
Two solid lines represents the relation between the spheroid mass and
the black hole mass in the local universe with the Eddington ratio 
($\lambda=1$). The upper and lower lines represent
the stellar mass estimations with $0.8\times10^{10} M_{\odot}$ 
and $2.0\times10^{10} M_{\odot}$ for $M_{\rm AB 1700}=-21$ mag,
respectively. See text for the details.
\label{host_Lx}}
\end{center}
\end{figure*}

Using the upper and lower limits of the brightnesses of the host galaxy
in the F606W and F775W bands, we derive the 
limits on their UV absolute magnitudes, $M_{\rm AB 1700}$. 
We interpolate or extrapolate the brightness limits in the two bands
to derive the limits at object-frame 1700{\AA}. The derived absolute magnitude 
limits are shown in Figure~\ref{host_Lx} against $L_{\rm 2-10keV}$. 
The upper limits of the ``compact'' AGNs are
distributed from $M_{\rm AB 1700} = -19.0$ to 
$M_{\rm AB 1700}=-21.5$ (left panel).
The lower and upper limits of the ``extended'' AGNs are
distributed from $M_{\rm AB 1700}=-17.0$ to 
$M_{\rm AB 1700}=-22.5$ (right panel). 
The knee of the UV luminosity function ($L^{*}$) of LBGs 
at $z\sim3$ is $M_{\rm AB 1700} = -21.2$ mag \citep{ste99},
thus the limits on the host galaxies correspond to 
$0.02L^* - 3L^{*}$. There is no clear difference between 
the upper limits on the luminosities of the host galaxies of
the ``compact'' and ``extended'' AGNs.

If we assume that star formation in the 
host galaxies is continuous at the time
of the observation, their UV luminosities reflect the star formation 
rates of the host galaxies. The star formation rate 
can be derived from UV luminosity with 
$SFR = 1.3(2.9)\times10^{-28} L_{\nu} (\lambda 1500)$
erg s$^{-1}$ Hz$^{-1}$ \citep{mad98} with Salpeter
(Scalo) initial mass function (IMF). 
The derived SFR with Salpeter IMF without extinction correction
is shown in the top horizontal axis of Figure~\ref{host_Lx}.
The limits of the UV absolute magnitudes of the host galaxies correspond to 
SFR of $0.3 - 40$ $M_{\odot}$ yr$^{-1}$.
The modest SFR is similar to the SFRs observed in the host galaxies of
optically-selected moderate luminosity QSOs 
at $1.8 < z < 2.75$ in the GEMS
field overlapping the GOODSS region (Jahnke et al. 2004).
 
\subsection{Stellar Mass of the Host Galaxies}

\subsubsection{Stellar Mass Estimated from UV Luminosity,
and the $M_{\rm bulge}-M_{\bullet}$ relation}

The stellar mass of a galaxy can be estimated with the
UV luminosity. There is a loose correlation between 
observed $M_{\rm AB 1700}$ and stellar mass
of $z\sim3$ LBGs estimated by a SED (Spectral Energy Distribution) 
fitting \citep{pap01}. 
For example an LBG with $M_{\rm AB 1700}=-21$ mag, its
stellar mass is estimated to be $0.8\times10^{10}\sim2.0\times10^{10}$ 
M$_{\odot}$ on average. The scatter of the mass estimation 
comes from the different IMF and metallicity used in the 
SED fitting. It should be noted that there is one order
of magnitude scatter in the stellar mass of 
each object from the average value. 

On the other hand, the black hole mass of the nucleus
can be estimated from the $L_{\rm 2-10keV}$ 
with assumptions about $L_{\rm 2-10keV}$ to bolometric 
luminosity  ($L_{\rm bol}$) conversion factor (bolometric correction) 
and the Eddington ratio, $\lambda$. The bolometric 
correction of typical non-obscured QSOs is calculated to be
$\log(L_{\rm bol}/L_{\rm 2-10keV}) = 1.54 +0.24L' +0.012L'^{2} -0.0015L'^{3}$,
where $L'=\log (L_{\rm bol}/10^{12})$ in unit of $L_{\odot}$ 
by Marconi et al. (2004).
With the bolometric correction and $\lambda$=1, 
an AGN with $L_{\rm 2-10keV}=10^{44}$
erg s$^{-1}$ is expected to have a bolometric luminosity of
$L_{\rm bol}=2.5\times10^{45}$ erg s$^{-1}$ and a black
hole mass of $2.7\times10^7$ M$_{\odot}$.

The estimation of the black hole mass from the continuum luminosity
of an AGN is plausible within a order of magnitude uncertainty. 
For optically-selected QSOs, there is a 
correlation between the continuum luminosity in the optical band and 
the black hole mass derived from the velocity width of a broad emission line
and the continuum luminosity
(Netzer 2003; Corbett et al. 2003; Warner, Hamann, \& Dietrich 2003) 
with a dispersion of about 1 dex. 
Considering that the SEDs of non-obscured QSOs are similar to each other,
we expect a correlation between the X-ray continuum luminosity
and the black hole mass.

Using the stellar mass and the black hole mass estimations, 
we plot the $M_{\rm bulge}-M_{\bullet}$ relation found
in the nearby galaxies
($M_{\bullet}/M_{\rm bulge}=0.014$; Haring \& Rix 2004)
on Figure~\ref{host_Lx} for a
case with $\lambda=1.0$. We assume that the estimated stellar mass
of the host galaxy represent that of the spheroidal component. 
The solid lines in Figure~\ref{host_Lx} 
represent the different stellar mass estimations
for the same UV luminosity ($0.8\sim2.0\times10^{10} M_{\odot}$
for $M_{\rm AB1700}=-21$mag).
Most of the upper limits on the luminosity of the host galaxy 
are distributed above the solid lines. 
This means, if we assume that the Eddington ratio does not
exceed one, the $M_{\rm bulge}-M_{\bullet}$ relation
in the local universe does not hold for the host galaxies of 
AGNs at high redshifts; the $M_{\rm bulge}$ of the host galaxies
of the AGNs at high redshifts are smaller than those of 
nearby galaxies with the same $M_{\bullet}$. 
However, the estimated stellar masses 
from UV-luminosity is insecure, because the UV luminosity
is dominated with bright early-type
stars with small $M/L$ ratio, and easily affected by dust extinction
($A_{\rm 1700} = 9.65 E(B-V)$; Calzetti et al. 2000). 

\subsubsection{Stellar Mass Estimation with K-band Magnitude for the 
"Extended" AGNs}

The stellar mass of the host galaxy of the "extended" AGN is 
constrained better with its $K_{\rm AB}$ magnitude, which corresponds 
to the object-frame optical wavelength, than based only on the 
UV-luminosity. The stellar mass of the host galaxy with the 
total $K_{\rm AB}$ magnitude is estimated as follows; 
first, we make models of galaxy SEDs with a stellar population 
synthesis code. Then we compare the models with the observed
SEDs of the host galaxy, changing the normalization of the 
models, which is equivalent to changing the stellar mass of the
model. We accept the SED models which do not exceed 
the upper and lower limits in all of the F606W, F775W, and 
F850LP bands. Finally, we select the SED models
which are fainter than the observed $K_{\rm AB}$ magnitude
from the accepted SED models, because the observed $K_{\rm AB}$ 
magnitude, which include the contribution from the nuclear 
component, is an upper limit on the host galaxy brightness.
The range of the stellar masses of the selected models is the
constraints on the stellar masses of the host galaxies.
For the ``compact'' AGNs, we can not constrain the lower
limits of the luminosity of the host galaxies, thus, the
stellar mass estimation is applicable to only the ``extended''
AGNs. Since $K_{AB}$-band magnitude of GOODSS089 is not 
available, we remove the object in discussions below.

The SED models of galaxies are made by using PEGASE.2 stellar
population synthesis code (Fioc \& Rocca-Volmerange 1997). 
The host galaxies of QSOs observed in the nearby universe tend to be 
early-type galaxies (e.g., Dunlop et al. 2003), thus we made models of E and
Sb type galaxies following the recipe in Kodama \& Arimoto (1997; $M_V=-22.49$)
and Le Borgne \& Rocca-Volmerange (2002), respectively.
The star formation histories of E and Sb models roughly
correspond to the instantaneous star formation and continuous
star formation models, respectively.

The E type galaxy model parameters are determined so as to 
reproduce the optical color-magnitude relation of
galaxies in the Coma cluster well (Kodama \& Arimoto 1997). 
The star formation efficiency of 3.33 Gyr$^{-1}$
and the infall time scale of 300 Myr are used. 
In the model, the galactic wind is introduced at
0.39 Gyr and stops the star formation at that time.
We use initial mass function (IMF) with a power of $-1.2$
with the mass range of 0.1 $M_{\odot}$ to 120 $M_{\odot}$.

The Sb type galaxy model parameters
are chosen to reproduce the optical colors of Sb galaxies 
in the nearby universe (Le Borgne \& Rocca-Volmerange 2002).
The star formation efficiency of 0.1 Gyr$^{-1}$
and the infall time scale of 8000 Myr are used. No galactic
wind is introduced. Salpeter IMF with mass range of
0.1 $M_{\odot}$ to 120 $M_{\odot}$ is assumed. 

The ages of the models are limited to be shorter than
those of the universe at the redshift of each AGN. We use 
only models with ages longer than 100 Myr. The age step of the 
calculations is about one tenth of each age. In both of the E and Sb 
galaxy models, initial and infalling gas 
contains no metal (0.0 $Z_{\odot}$). The metallicities
of the stars and the gases are consistently calculated. 
The metallicities of the gas at the 
age of 5 Gyr exceed the solar metallicity in both of the models.
The high metallicity of the gas component
of the QSOs even at $z=2\sim4$ 
(for review, Hamann \& Ferland 1999) is well reproduced with 
the E and Sb galaxy models.  
With the later-type galaxy 
recipes in Le Borgne \& Rocca-Volmerange (2002), we cannot 
reproduce the high-metallicity of the QSOs. 
For the internal extinction,
we do not use the model provided in the PEGASE.2 code, and 
apply the extinction curve for star-burst galaxies 
(Calzetti et al. 2000). We consider an extinction range 
of $E(B-V) = 0.0 - 1.0$ mag which covers the range of $E(B-V)$
estimated in $z\sim3$ LBGs ($0.0-0.5$; Papovich et al. 2001; 
Shapley et al. 2001) with step of 0.1 mag. 

\begin{figure*}
\begin{center}
\plotone{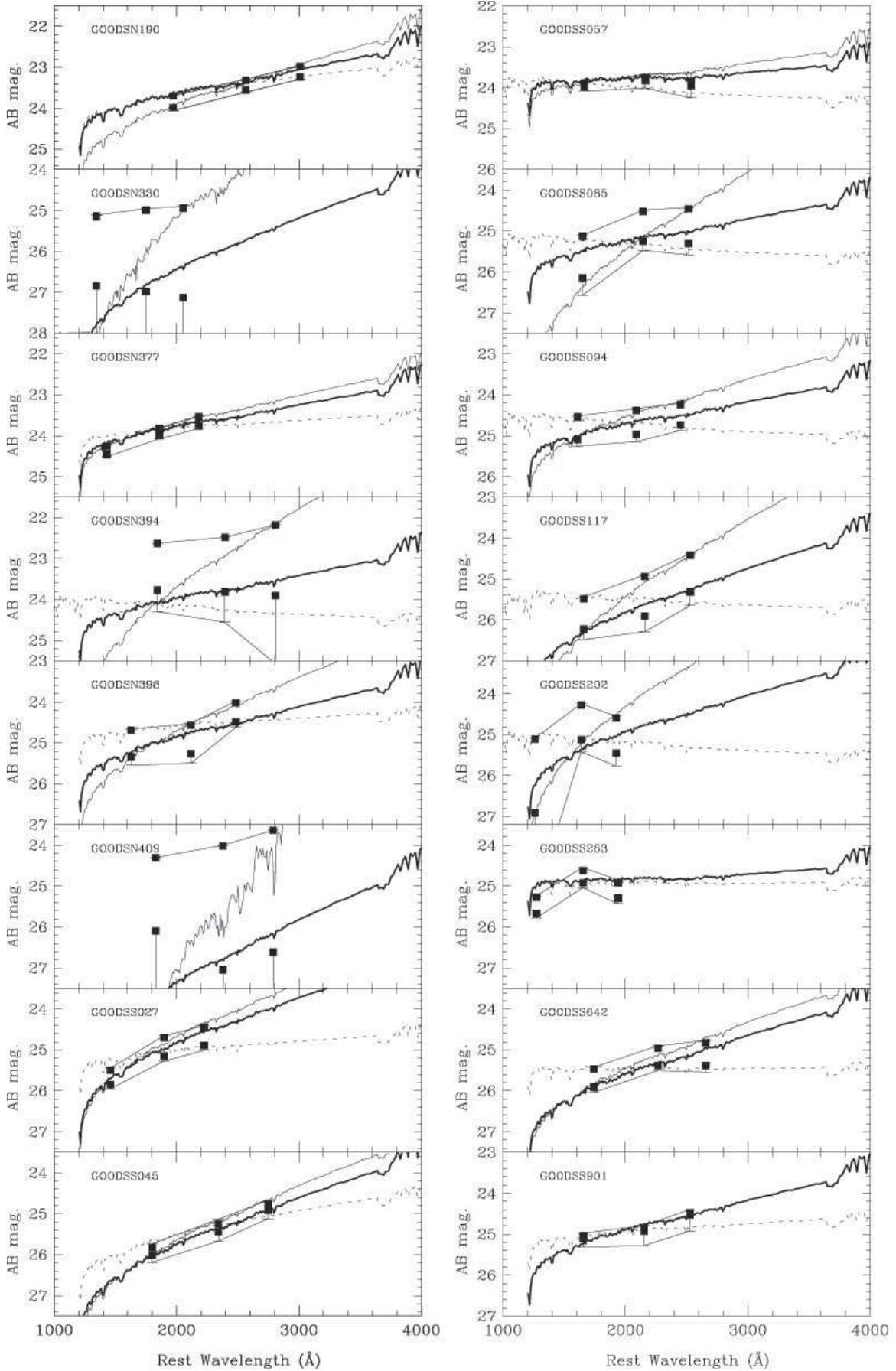}
\caption{
Upper and lower limits of the brightness
of the host galaxies of the ``extended'' AGNs
are shown with filled squares connected with solid lines,
along with the SEDs of the accepted models with
the largest (thin solid line) and smallest (thin dashed line)
stellar mass. The horizontal axis is the object-frame
wavelength and the vertical axis is the observed
AB magnitude. The models which do not exceed the
upper and the lower limits in the F606W, F775W, and F850LP
bands simultaneously with a certain normalization
are accepted. The thick solid line indicates the 
largest stellar mass model consistent with the
observed $K_{\rm AB}$-band magnitude.
\label{AGN_sed2}}
\end{center}
\end{figure*}

We convert the model SEDs to the observed AB magnitudes at
the redshift, using the filter and detector response functions of the ACS. 
We compare the model AB magnitudes with the upper and
lower limits in F606W, F775W, and F850LP-bands by changing 
the normalization of the model. We only accept a model
which do not exceed the upper limit, which is the 1$\sigma$ bright envelop
of the total magnitude, and the lower limit, which is the lower
envelop of the lower limits, 
all of the three bands simultaneously with a certain 
normalization. Because of the limited information, wide range
of the SED models are acceptable.
The example of the UV-constraints and the accepted 
models are shown in Figure~\ref{AGN_sed2}. In this figure, we 
plot the accepted models with the largest stellar mass 
(thin solid curves; usually they have redder color, old and dust 
reddened stellar population with large mass-to-light ratio) and the smallest 
stellar mass (thin dashed curves; usually they have bluer color, 
young stellar population with small mass-to-light ratio) for each 
AGN along with the derived upper and lower limits in the optical bands. 

\begin{figure*}
\begin{center}
\plotone{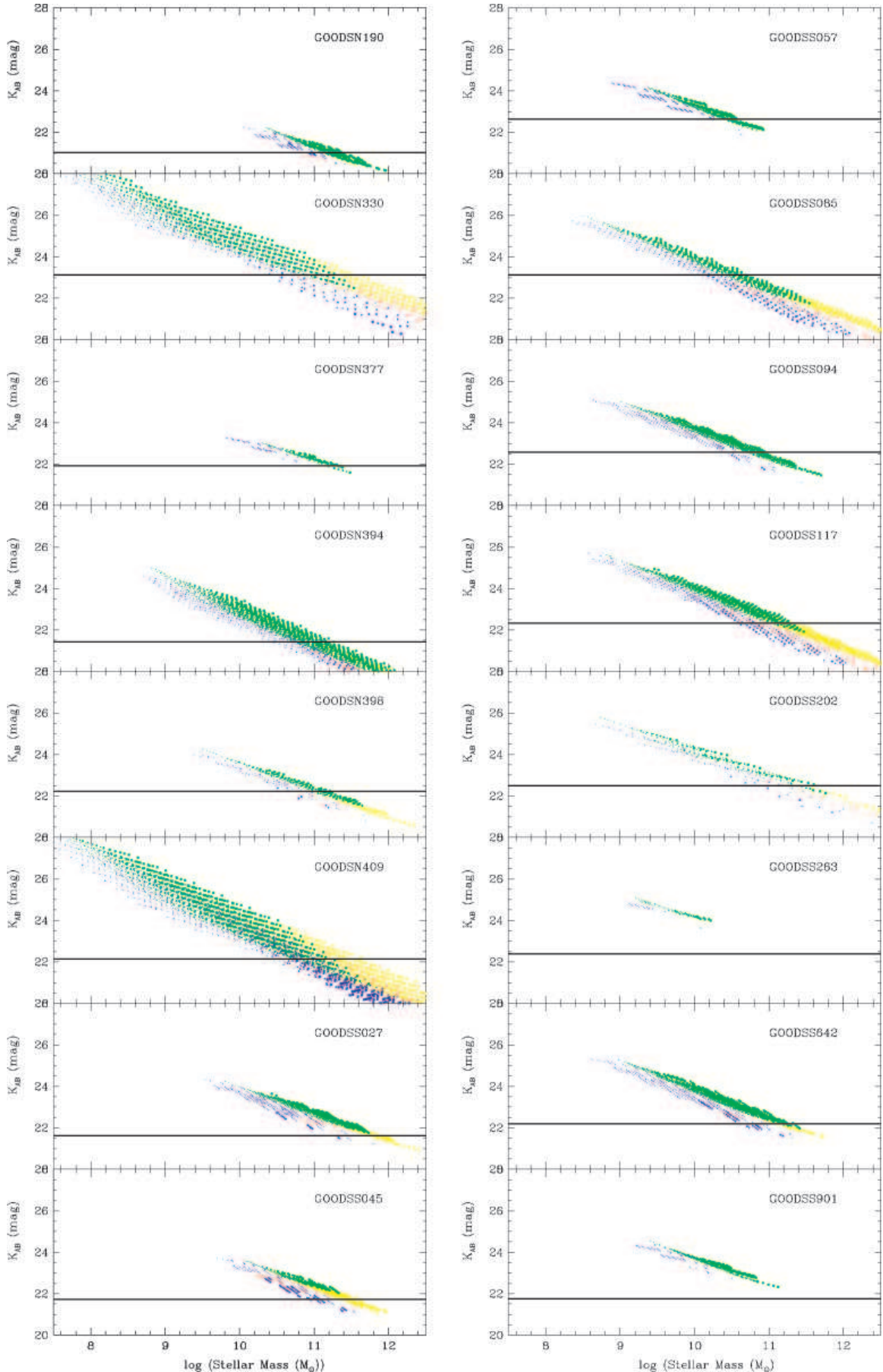}
\caption{Distribution of $K_{\rm AB}$-magnitudes
and stellar masses of the accepted models of each
AGN. Large, medium, and small marks corresponds to
models with age larger than 1000Myr, 
between 1000Myr and 500Myr, and smaller than
500 Myr, respectively. The red (yellow) and blue 
(green) points mean E (Sb) models with 
$E(B-V)>0.5$(mag) and $E(B-V)<0.5$(mag).
The horizontal thick solid line indicates the
observed $K_{\rm AB}$-magnitude of the AGN. 
\label{K_mass}}
\end{center}
\end{figure*}

For the accepted SED models and their normalizations,
we calculate the $K_{\rm AB}$-band magnitudes and 
stellar mass of the models. The distribution of 
the $K_{\rm AB}$ band magnitudes and the
stellar masses of the accepted models
for each object is shown in Figure~\ref{K_mass}. 
We select the accepted models whose $K_{\rm AB}$ band 
magnitudes are fainter than the observed $K_{\rm AB}$ magnitude.
Because the $K_{\rm AB}$ magnitude includes the nuclear 
component, the host galaxy magnitude need to be fainter than 
the $K_{\rm AB}$ magnitude, which is shown with a horizontal thick 
solid line. Basically, accepted models with 
brighter $K_{\rm AB}$-band magnitudes tend to have higher total mass. 
Thus the observed $K_{\rm AB}$-band magnitude 
constrains the upper total mass of the galaxy. The selected model
with the largest stellar mass is shown with the thick solid line
in Figure~\ref{AGN_sed2}.
Finally, the stellar mass of the host galaxy 
can be constrained from the mass range of the selected models.
The resulting stellar mass estimations are summarized in Table~\ref{Tab_mass}. 

\begin{figure}
\begin{center}
\plotone{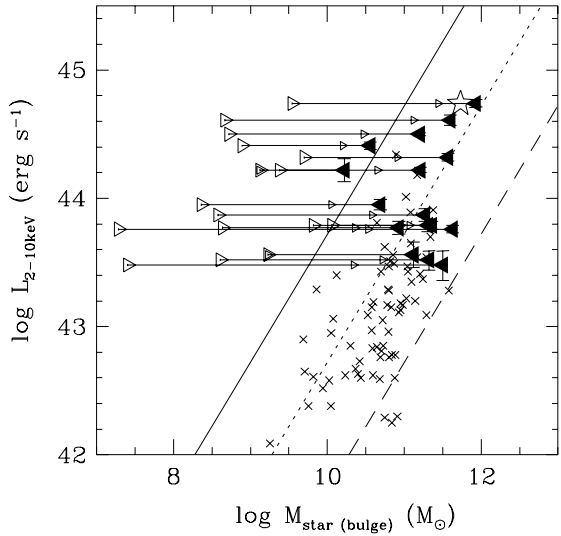}
\caption{
Distribution of stellar mass and
$L_{\rm 2-10keV}$ of the ``extended'' AGNs.
Upper and lower limits are shown with 
filled and open triangles, respectively. 
The small rightward open triangles indicate the
lower limits in case that the host galaxy
component dominates the whole $K_{AB}$-band magnitude.
Horizontal
solid lines indicate the allowed range of the stellar mass.
The star mark show the stellar mass of the best fit
model for GOODSN027 shown in Figure 11.
Crosses show the distribution of the bulge
stellar masses of the X-ray selected 
low-redshift AGNs from Schade et al. (2000). 
The solid line indicates $\lambda=1$, the 
bolometric correction of 25, and $M_{\bullet}/M_{\rm bulge}
=0.0014$. The dotted and dashed lines indicate the cases
with $\lambda = 0.1$ and 0.01, respectively.
\label{Lx_mass}}
\end{center}
\end{figure}

The resulting mass estimations are plotted in Figure~\ref{Lx_mass}
with triangles. The leftward filled triangles indicate the upper limits
on the stellar mass, the rightward open triangles indicate the lower
limits on the stellar mass. If we assume that the host galaxy
component dominates the whole $K_{AB}$-band magnitude, the lower
limit on the stellar mass becomes larger. The small rightward
open triangles indicate the lower limits. 
The $M_{\rm bulge}-M_{\bullet}$ relation in the 
nearby universe ($M_{\bullet}/M_{\rm bulge}=0.0014$; Haring \& Rix 2004) 
is also plotted with a solid (dotted, dashed) line for 
$\lambda=1$ ($\lambda=0.1$, $\lambda=0.01$) with the bolometric correction
same as that used in the previous section.
For luminous AGNs with $L_{\rm 2-10keV} > 10^{44}$
erg s$^{-1}$, the limits of the stellar masses of the host galaxies
are consistent with $\lambda=0.1$ or higher. For less luminous
AGNs, the limits are still consistent with $\lambda\sim0.05$.
All of the upper limits of the stellar masses, except for
GOODSS057 and GOODSS263, are consistent with $\lambda=1$,
which is different from Figure~\ref{host_Lx}. The difference indicates that
the mass to UV-light ratio of the accepted models with
the largest mass are larger than those of the LBGs at $z\sim3$
for the ``extended'' AGNs.

\subsubsection{Stellar Mass Estimation of GOODSS027}

\begin{deluxetable*}{llrlrlrlr}
\tabletypesize{\scriptsize}
\tablecaption{Limits on the Stellar Masses of the Host Galaxies of the ``Extended'' AGNs \label{Tab_mass}}
\tablewidth{0pt}
\tablehead{
\multicolumn{1}{c}{} &
\multicolumn{4}{c}{Accepted Model with the Largest Mass} &
\multicolumn{4}{c}{Accepted Model with the Smallest Mass} \\
\multicolumn{1}{c}{Name} &
\multicolumn{1}{c}{Type} &
\multicolumn{1}{c}{Age} &
\multicolumn{1}{c}{$E(B-V)$} &
\multicolumn{1}{c}{$\log$(Mass)} &
\multicolumn{1}{c}{Type} &
\multicolumn{1}{c}{Age} &
\multicolumn{1}{c}{$E(B-V)$} &
\multicolumn{1}{c}{$\log$(Mass)} \\
\multicolumn{1}{c}{} &
\multicolumn{1}{c}{} &
\multicolumn{1}{c}{(Myr)} &
\multicolumn{1}{c}{(mag)} &
\multicolumn{1}{c}{($M_{\odot}$)} &
\multicolumn{1}{c}{} &
\multicolumn{1}{c}{(Myr)} &
\multicolumn{1}{c}{(mag)} &
\multicolumn{1}{c}{($M_{\odot}$)} \\
}
\startdata
GOODSN190 & Sb & 3000 & 0.3 & 11.36 & E & 160 & 0.3 & 10.06 \\
GOODSN330 & Sb & 1000 & 0.7 & 11.50 & E & 120 & 0.0 &  7.40 \\
GOODSN377 & Sb & 1600 & 0.3 & 11.31 & E & 120 & 0.2 &  9.82 \\
GOODSN394 & Sb & 2500 & 0.3 & 11.20 & E & 120 & 0.0 &  8.72 \\
GOODSN398 & Sb & 2000 & 0.4 & 11.21 & E & 120 & 0.2 &  9.38 \\
GOODSN409 & Sb & 2500 & 0.8 & 11.63 & E & 120 & 0.0 &  7.28 \\
GOODSS027 & Sb & 1600 & 0.6 & 11.93 & E & 180 & 0.2 &  9.54 \\
GOODSS045 & Sb & 2500 & 0.6 & 11.57 & E & 140 & 0.4 &  9.70 \\
GOODSS057 & Sb & 2000 & 0.1 & 10.56 & E & 120 & 0.0 &  8.89 \\
GOODSS085 & Sb & 2000 & 0.3 & 10.69 & E & 120 & 0.0 &  8.36 \\
GOODSS094 & Sb & 2000 & 0.3 & 10.93 & E & 120 & 0.0 &  8.64 \\
GOODSS117 & Sb & 1200 & 0.6 & 11.27 & E & 200 & 0.0 &  8.58 \\
GOODSS202 & Sb & 1400 & 0.5 & 11.60 & E & 120 & 0.0 &  8.67 \\
GOODSS263 & Sb & 1400 & 0.1 & 10.22 & E & 120 & 0.1 &  9.12 \\
GOODSS642 & Sb & 2500 & 0.5 & 11.32 & E & 120 & 0.1 &  8.61 \\
GOODSS901 & Sb & 1800 & 0.4 & 11.12 & E & 120 & 0.2 &  9.23 \\
\enddata
\end{deluxetable*}

A part of the GOODSS field is covered by the ACS and the NICMOS
Ultra Deep Field (UDF). Using the deep optical and near-infrared data,
we obtain better constraint of stellar masses
for AGNs in the region. Four high-z AGNs locate in the ACS 
UDF, and two of them, GOODSS027 and GOODSS091 
are covered by the NICMOS UDF. GOODSS027 is mostly dominated
by the host galaxy component. The radial profile of the galaxy is well 
fitted with the de Vaucouleurs profile without a nuclear stellar component in 
the ultra deep image.
On the contrary, GOODSS091 image is dominated by a stellar component. 
By assuming that the contribution from a stellar nucleus is 
negligible for GOODSS027, we use the F110W and F160W magnitudes 
along with F606W, F775W, F850LP, and the ISAAC 
$K_{\rm S}$-band magnitudes to obtain a better constraint on 
the stellar mass of the object.

\begin{figure}
\begin{center}
\plotone{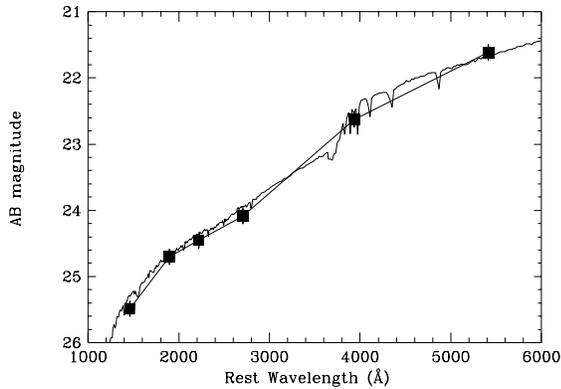}
\caption{
F606W to $K_s$-band SED of GOODSS27
shown with filled squares and error
bars along with the best fit model SED.
The horizontal axis is the object-frame wavelength
and the vertical axis is the observed AB magnitude.
\label{GOODSS27_sed}}
\end{center}
\end{figure}

In order to conduct SED fitting, we derived aperture
magnitudes in the F606W, F775W, F850LP, F110W, and F160W 
bands by adjusting the PSF size in all of the 5 images 
to that in the F160W band (FWHM of $0.^{\prime\prime}38$).
We used a $2.^{\prime\prime}6$ diameter to derive the
SED of the object. In order to convert the aperture magnitudes
to total magnitudes used in the $K$-band,
we add the difference between the aperture and total magnitude
in the F775W band to the aperture magnitudes. We fit the photometric 
data points with the stellar synthesis model spectra for E and Sb derived 
in the previous section. For the object, we use $\chi^2$
minimization method to derive the best fit synthesis model. 
The best fit model is shown in Figure~\ref{GOODSS27_sed}.
It is the Sb model at 1800 Myr with $E(B-V)=0.5$ mag and
a stellar mass of $5.3\times10^{11} M_{\odot}$. The model is
similar to the largest mass model of the object in the previous subsection.
The mass of the best fit model is shown with the star mark in 
Figure~\ref{Lx_mass}. The stellar mass of the model is 
close to the largest mass model of the object in the previous
subsection. The result is reasonable, because the upper limit of the
mass in the previous section is defined by the spectral model
that follows the total magnitudes in the F660W, F775W, F850LP,
and $K_{\rm AB}$ bands.
The derived stellar mass is larger than the upper limits on the
stellar masses of the other ``extended'' AGNs.

\section{Discussion}

\subsection{Comparison with Properties of Galaxies at $2<z<4$}

The object-frame UV absolute magnitudes of the host 
galaxies of the X-ray-selected AGNs at $2<z_{\rm sp}<4$ are 
distributed within a order of magnitude of the typical 
(knee of the luminosity function) UV absolute magnitude 
of the LBGs at $z\sim3$. The inferred star formation 
rates are less than 40 $M_{\odot}$ yr$^{-1}$ for most of them, 
also which are similar to those of LBGs. The constraints on the 
star-formation rates of the host galaxies reject 
apparently large ($>100 M_{\odot}$ yr$^{-1}$) star 
formation rates which are observed in high-z ultra-luminous
QSOs, if we neglect the effect of dust extinction.
The observed SEDs of the ``extended'' AGNs  
allow the synthesis models with a young stellar population 
affected by $E(B-V)=0.5$ mag, which corresponds to 
$A_{\rm 1700}=4.8$ mag with the Calzetti extinction curve. 
This means the reddening correction of up to
2 orders of magnitude is still possible, and we cannot
reject the large star formation rate in the high-z 
host galaxies. It should be noted that we cannot 
constrain violent star formation within 1kpc of the nucleus,
because such component cannot be distinguished from
the nuclear stellar component due to the resolution of
the observation.

The estimated upper limits of the stellar mass of the host galaxies of
the ``extended'' AGNs ($\sim 10^{10} M_{\odot} - 10^{11} M_{\odot}$) 
corresponds to the stellar mass of the $z\sim3$ LBGs with relatively
large stellar mass ($\sim 10^{9} M_{\odot} - 10^{11} M_{\odot}$; 
Papovich et al. 2001; Shapley et al. 2001), and similar to those of
red galaxies at $z\sim2$ ($2\times10^{10} M_{\odot} - 5\times10^{11}
M_{\odot}$ for Distant Red Galaxies, DRGs; F\"orster et al. 2004; van Dokkum et al. 2004). The estimated
upper limits are one order of magnitude smaller than radio galaxies,
which have the largest stellar mass at each redshift 
(Rocca-Volmerange et al. 2004). The lower limits of the stellar mass
are consistent with those of the LBGs at $z\sim3$ with small 
stellar mass.

\subsection{Comparison of the $M_{\rm bulge}-M_{\bullet}$ relation
with that of AGNs at Low-Redshifts}

In order to compare the $M_{\rm bulge}-L_{\rm 2-10keV}$ distribution 
of the ``extended'' AGNs with that of AGNs in the nearby universe, we also 
plot the bulge stellar masses and $L_{\rm 2-10keV}$ of Einstein Medium 
Sensitivity Survey (EMSS) sample of AGNs at $z<0.15$ from Schade et al. (2000)
with crosses. Their bulge $R_{\rm AB}$-band absolute magnitudes are
estimated based on their bulge apparent $I_{\rm AB}$-band magnitudes, 
which are determined from a snap-shot imaging survey with the HST and with
a color of an Sb galaxy at $z=0$ ($R_{\rm AB} - I_{\rm AB} = 0.65$ mag). 
To convert the $R_{\rm AB}-$band absolute magnitude to a bulge stellar mass, 
we use 
the average mass-to-light 
ratio of bulge component of early-type galaxies (4.1 in $R-$band; Haring \& Rix 2004). 
Their $L_{\rm 2-10keV}$ are derived from 0.5--3.5~keV flux 
measured in the EMSS survey by assuming a photon index of 1.8. 
The nearby AGN sample distribute between $\lambda = 0.01 - 1$,
and most of them are concentrate between $\lambda = 0.01 - 0.1$.
The distribution is consistent with the similar plot with 
optical nuclear absolute magnitude and bulge absolute 
magnitude which suggests $\lambda < 0.1$
for most of them (Schade et al. 2000; Dunlop et al. 2003;
Jahnke, Kuhlbrodt, \& Wisotzki 2004). With nuclear
hard X-ray luminosity, Miyaji et al. (2004) also obtained
a similar result at intermediate redshift. However, even in the low
redshift universe, still the distribution is not well 
established (Sanchez et al. 2004).

The estimated stellar mass limits are similar to the
the estimated stellar masses of the spheroidal components of 
nearby AGNs with lower luminosity as shown in Figure~\ref{Lx_mass}.
If we assume the high-z AGNs have the same 
$\lambda$ as the nearby AGNs and follow the same $L_{\rm 2-10keV}$
and $M_{\bullet}$ relation, the stellar mass of the high-z
AGNs have to be smaller than that of a galaxy with the same
black hole mass in the nearby universe. This 
implies that the black hole grew faster than the stellar 
component in the host galaxy. Considering the $M_{\rm bulge}-M_{\bullet}$ 
relation in the nearby universe, the high-z host galaxies require to grow at least 
$3 - 10$ times, accumulate more than $10^{11} M_{\odot}$ in 
stellar mass, from $z=2\sim4$ to $z=0$ under the assumption
that the black hole already reaches its final mass at $z=0$. 
The look back time of redshift 2 (4) is 10.2 (12.0) Gyr, thus 
if star formation continues with the upper limits of the inferred 
star formation rates ($<40 M_{\odot}$ yr$^{-1}$) in the host galaxies, 
it is still possible to gain $10^{11}$ $M_{\odot}$ of stars.
However, the black holes also can grow up since $z=2-4$.
An estimated average black hole growth history of black holes with
final $z=3$ mass of about $10^{8} M_{\odot}$ 
indicates that the mass at $z=2\sim4$ is 
only $10\sim30$\% in comparison with 
the final $z=0$ mass (Marconi et al. 2003). 

Alternatively, it is also possible that
the high-z AGNs have larger $\lambda$ value
than the nearby AGNs. The black hole mass estimations of 
the optically-selected luminous QSOs derived from the width of the 
optical broad emission line and the absolute magnitude suggest 
that the Eddington ratio
reaches 1 for a significant fraction of luminous AGNs (Woo \& Urry 2002;
Vestergaard 2004; McLure \& Dunlop 2003; Warner, Hamann, \& Dietrich 2004).
For example, 27\% of the AGNs have $\lambda$ larger than 1 
in Warner et al. (2004). The fraction gets higher for AGNs with larger
luminosity. For low-luminosity AGNs, the estimated Eddington 
ratios are mostly distributed between 0.01 and 0.1 (e.g., Schade et al. 2000), 
systematically smaller than those observed in luminous AGNs
(McLure \& Dunlop 2003). 

In order to distinguish the two possibilities, 
it is necessary to determine the mass of the central
black hole of high-redshift AGNs by an independent way.
For non-obscured ``compact'' AGNs, using the CIV line width
($3000 - 5000$ ks s$^{-1}$ for GOODSS AGNs; Szokoly et al. 2003) 
and the total $M_{\rm 1700 AB}$, we can estimate
the central black hole mass (e.q. (4) in Warner, Hamann, \& Dietrich 2003).
The estimated masses of the black holes are $1\sim4 \times 10^{8} M_{\odot}$ 
for $L_{\rm 2-10keV} = 10^{45}$ erg s$^{-1}$ and 
$1\sim4 \times 10^{7} M_{\odot}$ for $L_{\rm 2-10keV} = 10^{44}$ 
erg s$^{-1}$ AGNs. It is assumed that the absolute magnitude is not 
affected by the extinction. These masses indicate 
$\lambda = 0.3 - 1.3$, which are larger than those of 
low-luminosity AGNs in the nearby universe and similar 
to high-luminosity AGNs. 
If the ``extended'' AGNs have the same $\lambda$ for the ``compact'' AGNs,
all of the mass limits of the host galaxies are consistent with 
the $M_{\rm bulge}-M_{\bullet}$ relation in the nearby universe.

The estimated stellar mass ranges of the high-redshift host 
galaxies do have uncertainty of one order of magnitude 
even for the ``extended'' AGNs. 
In order to reduce the uncertainty, better 
constraints on the host galaxy apparent $K$-(or $H$-)
band magnitude is crucial, and additional $J$-band information
is helpful. Ground-based AO system will provide us 
unique high-angular resolution images in the 
near-infrared bands (e.g., Croom et al. 2004). 
Currently observable sample is restricted, 
only to relatively bright QSOs selected from the
huge sample of QSOs found in shallow wide area survey 
(like 2Qz, 2dF QSO survey), due to lack of natural 
guide star. The limitation
will be very much eased with laser guide star system 
in the near future. Moreover, the improvements of 
AO system with higher frequency corrections enable us 
to obtain high-resolution images in the shorter
wavelength range, such as $J$-band, constantly.

\acknowledgments

The author would like to thank Drs. Kouji Ohta, Kazuhiro Sekiguchi, 
Tadafumi Takata, Yoshihiro Ueda, and Kentaro Aoki for their detailed 
and invaluable comments. M.A. acknowledge support from a Research
Fellowship of the Japan Society for the 
Promotion of Science for Young Scientists.

\end{document}